\documentclass[pdflatex,sn-mathphys-ay]{sn-jnl}     

\usepackage{graphicx}                   %
\usepackage{multirow}                   %
\usepackage{amsmath,amssymb,amsfonts}   %
\usepackage{amsthm}                     %
\usepackage{mathrsfs}                   %
\usepackage[title]{appendix}            %
\usepackage{xcolor}                     %
\usepackage{textcomp}                   %
\usepackage{manyfoot}                   %
\usepackage{booktabs}                   %
\usepackage{algorithm}                  %
\usepackage{algorithmicx}               %
\usepackage{algpseudocode}              %
\usepackage{listings}                   %
\usepackage{makecell}                   %

\usepackage{lineno} 

\theoremstyle{thmstyleone}%
\theoremstyle{thmstyletwo}%

\theoremstyle{thmstylethree}%

\begin{document}

\title[Article Title]{Assessing Meteo-HySEA Performance for Adriatic Meteotsunami Events}


\title[Article Title]{Assessing Meteo-HySEA Performance for Adriatic Meteotsunami Events}


\author*[1]{\fnm{Alejandro} \sur{González}}\email{alexgp@uma.es}

\author[2,3]{\fnm{Cléa} \sur{Denamiel}}\email{clea.lumina.denamiel@irb.hr}
\equalcont{These authors contributed equally to this work.}

\author[1]{\fnm{Jorge} \sur{Macías}}\email{jmacias@uma.es}
\equalcont{These authors contributed equally to this work.}

\affil*[1]{\orgdiv{Departamento de Análisis Matemático, Estadística e Investigación Operativa y Matemática Aplicada}, \orgname{University of Málaga}, \orgaddress{\street{Campus de Teatinos}, \city{Málaga}, \postcode{29071},  \country{Spain}}}

\affil[2]{\orgdiv{Division for Marine and Environmental Research}, \orgname{Ruder Bošković Institute}, \orgaddress{\street{Bijenička cesta 54}, \city{Zagreb}, \postcode{10000}, \country{Croatia}}}

\affil[3]{\orgname{Institute for Adriatic Crops and Karst Reclamation}, \orgaddress{\street{Put Duilova 11}, \city{Split}, \postcode{21000}, \country{Croatia}}}


\abstract{Meteotsunamis are atmospherically driven sea-level oscillations that can trigger hazardous coastal flooding, particularly in resonant bays. This study assesses the GPU-based Meteo-HySEA model for meteotsunami simulation in the Adriatic Sea, benchmarking its performance against the CPU-based AdriSC-ADCIRC system. Three documented events (2014, 2017, 2020) were simulated using WRF downscaling of ERA reanalyses and validated with tide-gauge and microbarograph observations. Both models are limited by the underestimation of mesoscale pressure disturbances in the atmospheric forcing. Meteo-HySEA generally reproduces the timing and spatial variability of sea-level oscillations and often yields larger amplitudes than ADCIRC, but it tends to overestimate dominant wave periods, particularly in enclosed basins. Differences in oscillation persistence underscore the need for further validation against high-resolution tide-gauge data to assess whether Meteo-HySEA captures harbor seiches more realistically or ADCIRC better represents physical energy dissipation. Crucially, GPU acceleration provides order-of-magnitude gains in computational efficiency, enabling rapid high-resolution, multi-grid simulations including inundation, and thus offering strong potential for operational early warning.}

\keywords{Meteo-HySEA, GPU, Meteotsunami, Adriatic Sea, AdriSC, ADCIRC}



\maketitle

\section{Introduction}\label{sec:introduction}

Meteotsunamis—tsunami-like sea-level oscillations generated by atmospheric disturbances \citep{Monserrat2006}—are a well-documented hazard in various parts of the world (e.g., \citealt{Zemunik2022}). These high-frequency sea-level events can have devastating consequences for coastal infrastructure and human safety, particularly in narrow bays and harbors where resonance effects can significantly amplify wave heights \citep{Salaree2018}. Unlike seismic tsunamis, which are more widely recognized by both scientists and the public, meteotsunamis often occur with little warning and are frequently underestimated in risk assessments \citep{Lewis2024}. Yet, recent destructive events in the Mediterranean \citep{Orlic2010}, the Great Lakes \citep{Bechle2016}, and other regions have highlighted that their impacts can be comparable to those of “classical” tsunamis, particularly when they coincide with high tides or storm surges \citep{Tojcic2021}. The accurate simulation and forecasting of meteotsunamis are, therefore, critical for both operational early warning systems and fundamental research into their dynamics (e.g., \citealt{Renault2011,Angove2021,AndersonMann2021,SunNiu2021,Rahimian2022,Kim2022}).

\begin{figure}
    \centering
    \includegraphics[width=0.8\linewidth]{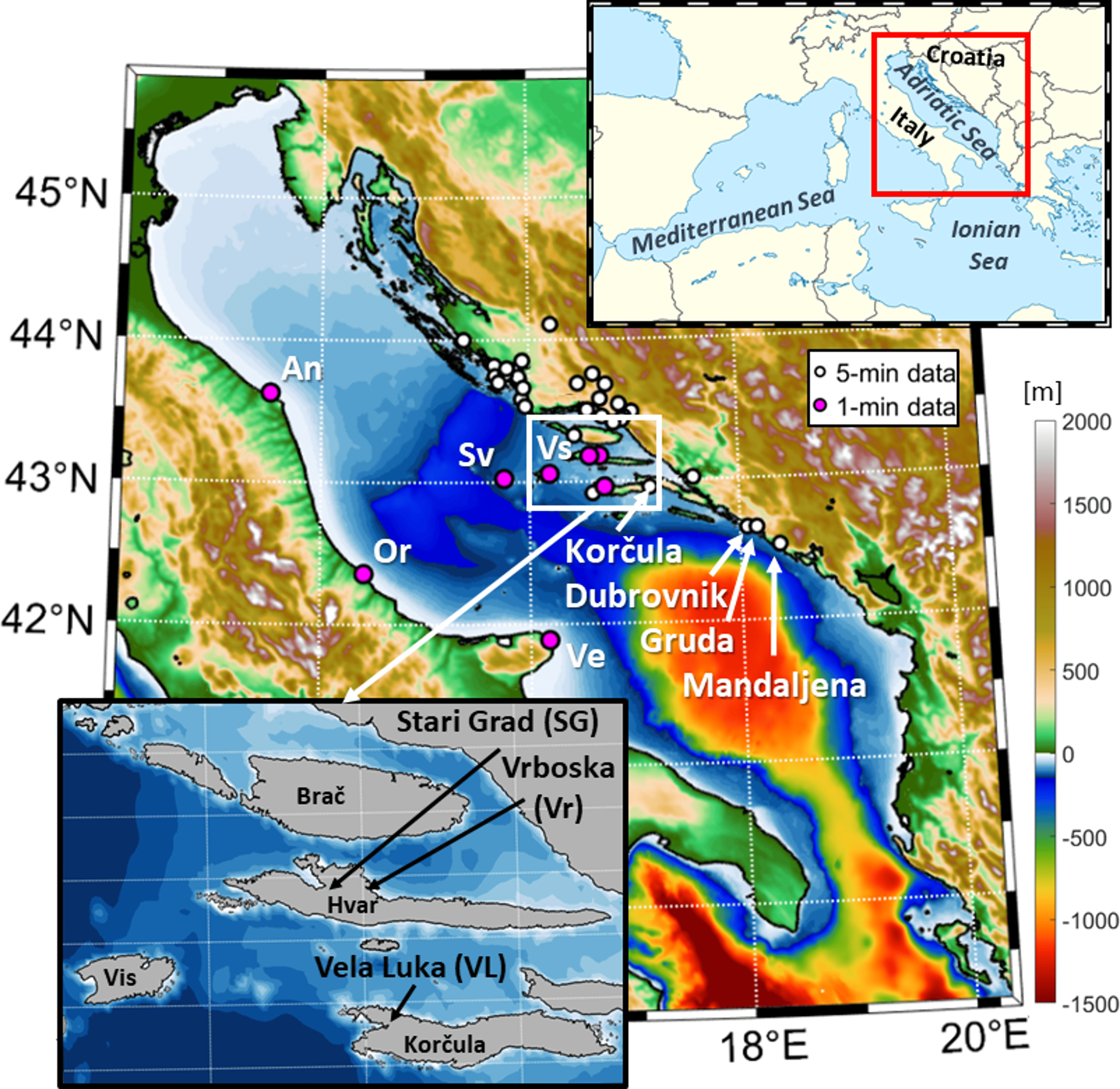}
    \caption{Location and topo-bathymetry of the Adriatic basin with a zoom on the study area including the Vela Luka, Stari Grad and Vrboska harbors. Observational network based on 1-min data at microbarograph locations (pink circles) and at tide-gauge locations in the Vela Luka and Stari Grad harbors as well as 5-min data from various pressure sensors (white dots).}
    \label{fig:fig01}
\end{figure}

In the Adriatic Sea (Fig. \ref{fig:fig01}), the Adriatic Sea and Coast (AdriSC) modeling suite has been successfully employed to simulate meteotsunamis using Central Processing Unit (CPU)-based high-resolution atmospheric and oceanic models—particularly in vulnerable bays and harbors such as Vela Luka, Stari Grad, and Vrboska. The AdriSC system has demonstrated strong capability in reconstructing historical meteotsunami events, contributing both to early warning strategies and to advancing understanding of atmosphere–ocean interactions specific to the region \citep{Denamiel2018,Denamiel2019a,Denamiel2019b,Denamiel2022a,Tojcic2021}. These successes highlight the Adriatic as a natural laboratory for investigating the physics of meteotsunamis and for testing forecasting strategies that could later be transferred to other regions.

However, the computational cost of CPU-based simulations like those in AdriSC remains a critical bottleneck—particularly for ensemble forecasting, probabilistic hazard assessment, or real-time early warning. Operational centers require near-real-time results, and the difference between simulations taking several hours versus just minutes can dictate the effectiveness of warning systems (e.g., a 6-hour tsunami forecast took only 1.5 minutes on a GPU). Beyond speed, parallel GPU processing enables multiple simulations in parallel, making sensitivity and uncertainty analyses more feasible. For example, the GPU-augmented GeoClaw model achieved a 3.6–6.4× speedup over a 16-core CPU \citep{Qin2019}. The Volna-OP2 model further demonstrates this capability: a 6-h high-resolution inundation run completed in just 2.25 minutes on a single Nvidia Tesla V100 GPU \citep{Giles2021}. Broadly, the computational efficiency and energy performance of GPU architectures have driven their adoption in operational ocean forecasting systems worldwide \citep{PorterHeimbach2025} as well as in tsunami simulations (e.g., \citealt{Macias2020b}).

In this study, we present and evaluate the latest developments of the new HySEA model adapted to simulate meteotsunamis (hereafter Meteo-HySEA). Meteo-HySEA belongs to the HySEA model family developed by the EDANYA research group at the University of Málaga, Spain. This family comprises numerical models for tsunami propagation and inundation. These models implement state-of-the-art finite volume methods, combining robustness, reliability, and accuracy within a GPU-based framework that enables simulations faster than real time. Tsunami-HySEA has undergone extensive testing \citep{Castro2005, CASTRO2006, Diaz2008, Asuncion2013, Lynett2017} and has been validated and verified according to the standards of the U.S. National Tsunami Hazard Mitigation Program (NTHMP) \citep{Macias2017, Macias2020a, Macias2020b}. Both Tsunami-HySEA and Landslide-HySEA, the code for landslide-generated tsunamis, are recognized as flagship European codes within major projects such as ChEESE, ChEESE-2P, DT-GEO, Geo-INQUIRE, and eFlows4HPC. Tsunami-HySEA is also operational in several tsunami early warning systems, including those of the Spanish National Geographic Institute (IGN), the Hydrographic and Oceanographic Service of the Chilean Navy (SHOA), the Pacific Marine Environmental Laboratory of the National Oceanic and Atmospheric Administration (PMEL-NOAA, U.S.), and the National Institute of Geophysics and Volcanology (INGV, Italy).

Meteo-HySEA is the new GPU-based framework that incorporates high-resolution atmospheric pressure fields as forcing and is capable of simulating meteotsunami generation, propagation, coastal amplification, and inundation in a computationally efficient manner. By integrating inundation capabilities, Meteo-HySEA goes beyond open-water propagation to address one of the most critical aspects of preparedness: the onshore impact of meteotsunamis on communities and infrastructure. Early functionalities of Meteo-HySEA were showcased
in \cite{Cheese2P}. The software has also been validated against laboratory experiments in reproducing Proudman resonance (e.g., \citealt{Vilibic2008, inproceedingsBeisiegel}), as well as in a real-world test case in the Gulf of Mexico in 2010 \citep{NTHMP2022}, employing actual bathymetric data and synthetic atmospheric pressure forcing \citep{2024EGUGA..26.8317G}. Furthermore, in \cite{GonzalezPino2025} the numerical scheme is detailed and some of these results are presented.

Given the demonstrated strengths of AdriSC in the Adriatic Sea, we propose to test the performance of Meteo-HySEA by applying it to a set of well-documented historical meteotsunami events in the region, specifically the events of June 2014, June–July 2017, and May 2020. These events are characterized by high-quality observational datasets and/or previous successful AdriSC simulations, making them ideal case studies for validating the new modeling system.

Through this model intercomparison approach, we aim to assess the skill of the Meteo-HySEA model in capturing key features of observed meteotsunamis and to identify potential advantages and limitations relative to traditional CPU-based modeling frameworks. Beyond academic interest, our findings will provide insights into how GPU-based models can be operationalized within early warning systems, where computational speed, reliability, and inundation detail are essential for timely risk reduction.

The article is structured as follows. Section \ref{sec:sec2} introduces the models, datasets, and methods used in this study. Section \ref{sec:sec3} presents a comparison between AdriSC and Meteo-HySEA simulations for three well-documented Adriatic meteotsunami events (June 2014, June–July 2017, and May 2020). The performance of Meteo-HySEA in reproducing meteotsunamis in the Adriatic Sea is further analyzed in Section \ref{sec:sec4}, followed by the main conclusions in Section \ref{sec:conclusions}.

\section{Models, Data and Methods}\label{sec:sec2}

\subsection{Adriatic Sea and Coast (AdriSC) modeling suite}\label{sec:sec2.2}

In the context of this study, the AdriSC modeling framework comprises two interconnected components: a basic module designed to simulate kilometer-scale atmospheric and oceanic dynamics across the broader Adriatic region \citep{Denamiel2021a,Denamiel2022b,Pranic2021}, and a specialized meteotsunami module dedicated to high-resolution hazard assessment \citep{Denamiel2019a}.

\begin{figure}
    \centering
    \includegraphics[width=0.8\linewidth]{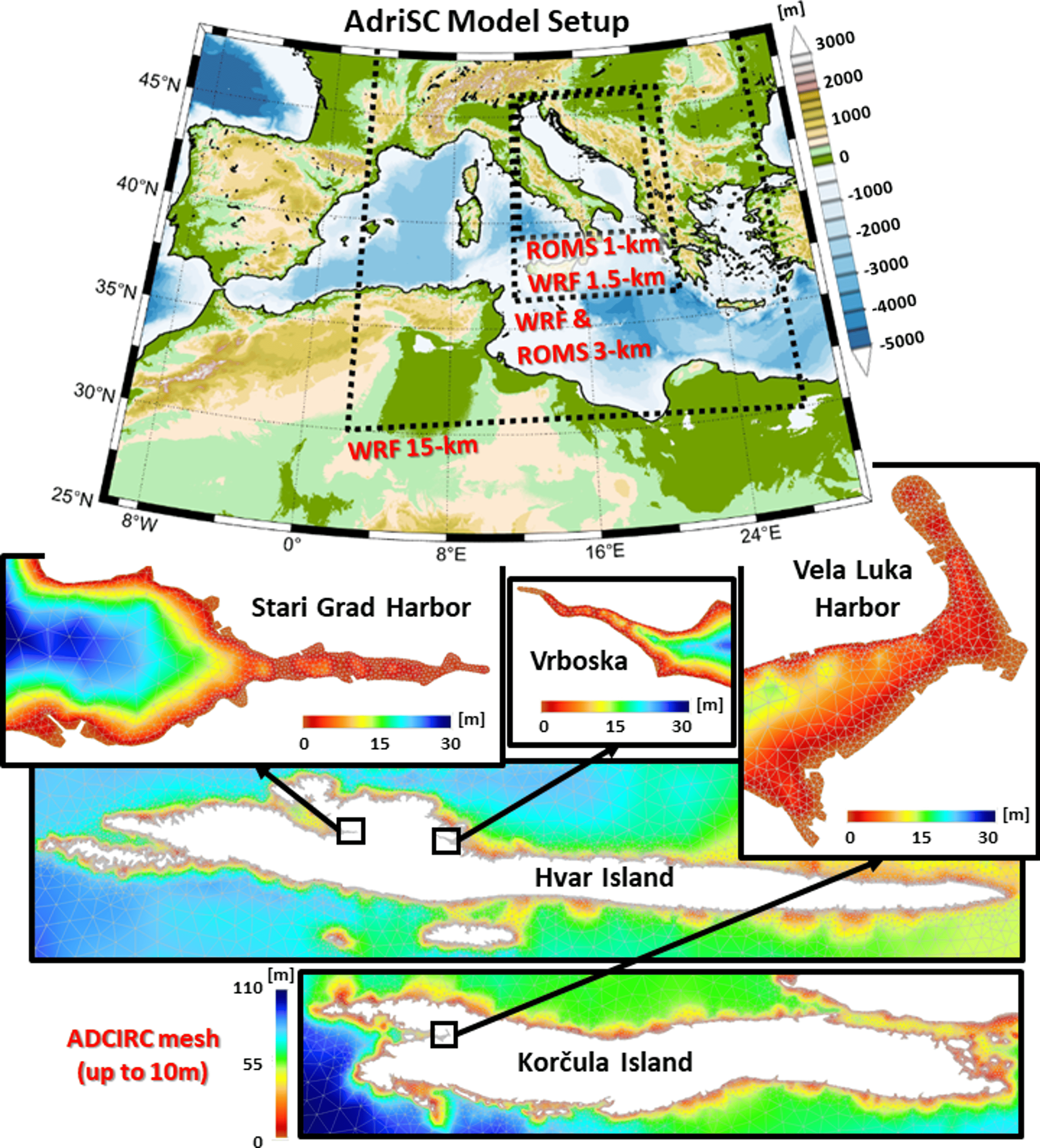}
    \caption{Adriatic Sea and Coast (AdriSC) modeling suite setup including the domains of the different models used (WRF, ROMS and ADCIRC) with a zoom on the ADCIRC model mesh at the locations of interest: Vela Luka, Stari Grad and Vrboska harbors.}
    \label{fig:fig02}
\end{figure}

At its core, the basic module is based on an enhanced implementation of the Coupled Ocean–Atmosphere–Wave–Sediment–Transport (COAWST) system \citep{Warner2010}. This system leverages the Model Coupling Toolkit (MCT) to achieve dynamic coupling among the Weather Research and Forecasting (WRF; \citealt{Skamarock2004}) atmospheric model and the Regional Ocean Modeling System (ROMS; \citealt{Shchepetkin2009}). Within this configuration, WRF is executed on two nested domains at 15 km and 3 km resolution, spanning the central Mediterranean and the Adriatic–Ionian region, respectively (Fig. \ref{fig:fig02}). Correspondingly, ROMS operates on 3 km and 1 km resolution grids, covering the Adriatic–Ionian domain and the Adriatic basin (Fig. \ref{fig:fig02}).

 On the other hand, the dedicated meteotsunami component employs an offline coupling strategy wherein the WRF model, downscaled to a 1.5 km resolution over the Adriatic, drives the two-dimensional depth-integrated (2DDI) ADvanced CIRCulation (hereafter, AdriSC-ADCIRC) model on an unstructured mesh refined to spatial resolutions of up to 10 m in zones particularly prone to meteotsunami impact (e.g., Vela Luka, Stari Grad and Vrboska; Fig. \ref{fig:fig02}). The coupling workflow begins with hourly atmospheric outputs from the 3 km WRF domain, which are further downscaled to the 1.5 km domain and saved every minute. These WRF outputs are then used in conjunction with hourly sea-surface elevations from the ROMS 1 km grid to force the AdriSC-ADCIRC model. In this configuration, the ADCIRC model thus receives minute-scale wind and pressure forcing from the WRF 1.5 km grid and hourly ocean boundary conditions—incorporating tidal contributions—from the ROMS simulations at the southern boundary near the Strait of Otranto.

All AdriSC simulations are performed over different durations adapted to the development of each meteotsunami multi-day events—specifically 4 days for the June 2014 events, 6 days for the June–July 2017 events, and 11 days for the May 2020 events. However, the first two days of each simulation are considered a spin-up period, allowing the model to stabilize, and only the remaining time is used for analysis. Additionally, the atmospheric and oceanic forcing fields used to force the AdriSC WRF and ROMS models are based on state-of-the-art reanalysis products. Specifically, the AdriSC WRF 15-km domain is initialized and forced using both the ERA-Interim (ERAI; \citealt{Dee2011}) and ERA5 \citep{Hersbach2020} reanalyses. However, ERAI was discontinued in August 2019 and therefore is not available for simulations beyond that date. As a result, only the ERA5 reanalysis was used to force the 2020 meteotsunami event simulations. Hereafter, WRF-ERAI and WRF-ERA5 refer to the 1-minute outputs from the AdriSC WRF 1.5-km simulations, which are downscaled from the AdriSC WRF 3-km model nested within the 15-km domain forced by the ERAI and ERA5 reanalyses, respectively. For the oceanic component, the ROMS 3-km domain is initialized and forced at the open boundaries with the Copernicus Mediterranean Sea reanalysis (MEDSEA; \citealt{Escudier2021}).

\subsection{Meteo-HySEA model}\label{sec:sec2.1}
Meteo-HySEA is a member of the HySEA model family, designed to provide the meteotsunami research community with advanced tools and expertise comparable to those available in other tsunami research fields, taking advantage of previous developments commonly used to accurately simulate tsunamis generated by various sources, such as earthquakes \citep{Macias2017, Macias2020a, Macias2020b} and landslides \citep{Asuncion2013a, Macias2015a, Macias2017a, GonzalezVida2019, EspostiOngaro2025}. Meteo-HySEA inherits two key features from the HySEA family: (i) a two-way nested grid algorithm and (ii) multi-GPU parallelization. The nested grid system supports high-resolution inundation modeling and the accurate simulation of tsunami-like waves in coastal regions, while the multi-GPU approach harnesses the computational power of multiple graphics cards to drastically reduce simulation times.

\noindent Meteo-HySEA reads time-dependent pressure fields defined on a structured grid, with the strict requirement that their spatial coordinates match those of the zero-level topobathymetric grid in nested simulations. Along the time dimension, the pressure forcing is interpolated within the model. This interpolation is essential due to the disparity between the tsunami solver’s time steps (typically $\Delta t = 10^{-1}, 10^{-2}, 10^{-3}, \dots$ seconds) and the much coarser temporal resolution of atmospheric models (with $\Delta t$ on the order of minutes). To balance accuracy and efficiency, Meteo-HySEA only loads into memory two consecutive atmospheric pressure snapshots—the ones bracketing the current simulation time—and applies convex linear interpolation based on the local tsunami time step.

\begin{table}[h]
\caption{Setup of the Meteo-HySEA nested grids for the Adriatic Sea. Each entry contains the extension covered by the grids and the number of cells of the grid.}\label{tab:tab01}
\begin{tabular}{@{}lcc@{}}
\toprule
Resolution & Grid Extension & Number of Cells \\
\toprule
1km  & $[12.07^{\circ},20.34^{\circ}]\times [39.26^{\circ}, 45.75^{\circ}]$ & $893\times 701$ \\
\midrule
250m  & $[13.44^{\circ},18.24^{\circ}]\times [41.59^{\circ}, 44.5^{\circ}]$ & $2072\times 1256$ \\
\midrule
60m  & $[16.16^{\circ},17.21^{\circ}]\times [42.67^{\circ}, 43.27^{\circ}]$ & $1808\times 1040$ \\
\midrule
& \multicolumn{1}{c}{Vela Luka} \\
\midrule
30m  & $[16.6^{\circ},16.74^{\circ}]\times [42.9^{\circ}, 43^{\circ}]$ & $492\times 360$  \\
\midrule
7m  & $[16.65^{\circ},16.72^{\circ}]\times [42.94^{\circ}, 42.99^{\circ}]$ & $972\times 680$ \\
\midrule
& \multicolumn{1}{c}{Stari Grad} \\
\midrule
30m  & $[16.48^{\circ},16.61^{\circ}]\times [43.17^{\circ}, 43.24^{\circ}]$ & $472\times 240$ \\
\midrule
7m  & $[16.51^{\circ},16.6^{\circ}]\times [43.17^{\circ}, 43.22^{\circ}]$ & $1228\times 700$ \\
\midrule
& \multicolumn{1}{c}{Vrboska} \\
\midrule
30m  & $[16.64^{\circ},16.8^{\circ}]\times [43.15^{\circ}, 43.23^{\circ}]$ & $568\times 274$ \\
\midrule
7m  & $[16.65^{\circ},16.72^{\circ}]\times [43.16^{\circ}, 43.21^{\circ}]$ & $944\times 688$ \\
\bottomrule
\end{tabular}
\end{table}

\begin{figure}
    \centering
    \includegraphics[width=.8\linewidth]{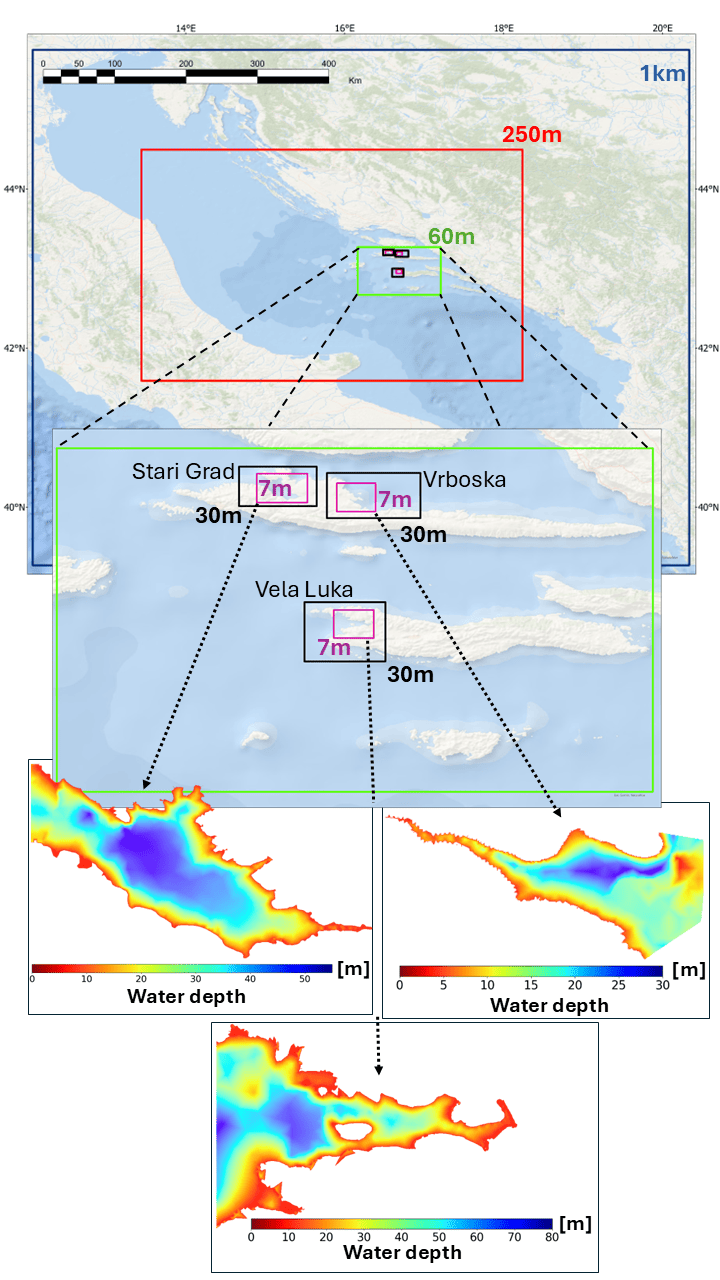}
    \caption{Meteo-HySEA nested grid system setup. A five-level nested grid system is shown, where the outlines indicate the extent of each grid. The spatial resolutions range from 1\,km (blue), 250\,m (red), 60\,m (green), 30\,m (black), and 7\,m (magenta). The bottom images display zoomed views of the highest-resolution grids covering the areas of Stari Grad, Vrboska, and Vela Luka.}
    \label{fig:fig03}
\end{figure}

The Meteo-HySEA model relies on pressure forcing from the WRF-ERAI and WRF-ERA5 models and uses initial sea level conditions derived from the AdriSC ROMS 1-km simulation, but does not include tidal or large-scale sea-level boundary forcing at the open boundary, focusing purely on the atmospheric disturbance-driven dynamics. Both the AdriSC-ADCIRC and Meteo-HySEA models are thus consistently forced by the WRF-ERAI and WRF-ERA5 simulations, enabling comparative evaluation of model skill under differing atmospheric reanalysis conditions, as previously demonstrated in \cite{Denamiel2019a, Denamiel2019b}.

The Meteo-HySEA simulation setup for the meteotsunami events analyzed in this study is illustrated in Fig. \ref{fig:fig03}. A five-level nested grid system is employed to achieve high-spatial resolution in the target locations. The configuration begins with a 1 km grid covering the entire Adriatic Sea, which nests into a 250 m grid. This grid, in turn, contains a 62.5 m grid (60 m approx.) encompassing the islands of Hvar and Korčula. For each specific site (Stari Grad, Vela Luka, and Vrboska), a 31.25 m grid (30 m approx.) nested within the 60 m grid is generated, and finally, a 7 m grid is used to resolve the bays and harbors in detail. Since the zero-level grid has a spatial resolution of 1 km, the atmospheric pressure fields from the 1.5 km WRF-ERAI and WRF-ERA5 models are resampled to match this resolution. The spatial extent and cell count of each grid used in the Meteo-HySEA simulations are summarized in Table \ref{tab:tab01}.

\subsection{Meteotsunami events}\label{sec:sec2.3}

The present study focuses on the detailed analysis of meteotsunami impacts within the harbors of Vela Luka (Korčula Island; Fig. \ref{fig:fig01}), Stari Grad, and Vrboska (both on Hvar Island; Fig. \ref{fig:fig01}) during three major events: 25–26 June 2014, 30 June–1 July 2017, and 11–19 May 2020. These locations were selected because they are among the most meteotsunami-prone sites along the Croatian coastline \citep{SepicOrlic2022}, frequently exhibiting strong resonant amplification due to their elongated, semi-enclosed geomorphology and their orientation relative to typical atmospheric disturbance tracks (Fig. \ref{fig:fig02}). Additionally, the selected events are among the most energetic and long-lasting meteotsunamis observed in the Adriatic Sea, with measurable impacts on coastal infrastructure, navigation, and harbor safety. Importantly, they were either fully or partially recorded by the observational network deployed during the MESSI project (Meteotsunamis in the Eastern Adriatic – Solving the mystery of high-frequency sea-level oscillations) described in \cite{Denamiel2019a, Denamiel2019b} which enabled high-resolution spatiotemporal observations of sea-level oscillations and atmospheric pressure disturbances. While the June 2014 event occurred prior to the full deployment of the MESSI instruments, it remains a reference case due to the exceptional performance of the AdriSC atmosphere-ocean modeling suite in reproducing its key characteristics, thereby demonstrating the robustness and reliability of the modeling framework for meteotsunami research.

On 25–26 June 2014, an exceptional multi-meteotsunami event impacted several locations along the Croatian coast with large oscillations recorded in Vela Luka and Stari Grad, where maximum sea-levels reached up to 1.5 m within the harbors. The event was triggered by a train of atmospheric gravity waves propagating from the Tyrrhenian Sea across the Adriatic, producing rapid pressure perturbations of up to 2.4 hPa in 5 minutes. These atmospheric disturbances propagated at speeds close to the local shallow-water wave celerity, creating conditions favorable for Proudman resonance (e.g., \citealt{Denamiel2019a, Denamiel2019b}). 

Other strong meteotsunami events occurred between 28 June and 1 July 2017, particularly affecting the bays of Stari Grad and Vrboska in Hvar Island. The oscillations lasted for nearly 24 hours, with maximum amplitudes up to 0.69 m recorded at tide-gauges and exceeding 1 m as observed in videos from the Vrboska harbor. These waves caused alternating sea withdrawal and sudden flooding, damaging small vessels and coastal infrastructure. The meteotsunamis were associated with atmospheric disturbances induced by a synoptic cyclone over the central Mediterranean, which generated upper-level jet stream winds exceeding 55 m/s at 500 hPa. Surface pressure perturbations of around 1–1.3 hPa occurred nearly simultaneously with the onset of the sea-level anomalies, supporting the theory of atmospheric resonance. Synoptic analyses and pressure records confirmed the presence of atmospheric gravity waves aligned with the resonance criteria for meteotsunami generation in the semi-enclosed bays of Hvar (e.g., \citealt{Denamiel2019a, Denamiel2019b}).

Finally, between 11 and 19 May 2020, a remarkable sequence of meteotsunami events occurred along the eastern Adriatic coast, particularly affecting the bays of Vela Luka, Stari Grad, and Vrboska. These sea-level oscillations—ranging from 0.6 to 0.8 m—were driven by recurring high-frequency atmospheric pressure disturbances (2–4 hPa over 5–15 min), combined with resonant responses in semi-enclosed bays \citep{Tojcic2021}. On 11 May, a significant event struck Vela Luka and Vrboska, where tide-gauge records indicated up to 0.81 m peak-to-trough waves of about 16 min period. Simultaneously, intense pressure oscillations were recorded by microbarographs—up to 3.9 hPa in Vela Luka and 2.5 hPa in Vrboska—during the early morning hours. Local eyewitnesses described harbor flooding, boats ashore, and high tidal variations across multiple days. Further events on 14–16 May produced repeated waves up to 0.8 m with periods of 12–18 min in the Vela Luka, Stari Grad and Vrboska harbors, with slightly weaker pressure dips (about 1.4 hPa). Media reports highlighted drying harbors and renewed flooding on multiple days, corroborating tide-gauge measurements. After 16 May, intermittent pressure waves persisted until 19 May, though sea-level fluctuations remained under 0.3 m, and no further flooding was observed.

\subsection{Observations}\label{sec:sec2.4}

The first set of observations used in this study originates from the MESSI observational network \citep{IZOR2025}, which was specifically designed to monitor meteotsunami-related atmospheric and sea-level disturbances across the Adriatic Sea. This network includes eight high-frequency microbarographs that measure atmospheric pressure at 1-minute intervals using Väisälä PTB330 sensors with a precision of ±0.01 hPa. In addition, two high-resolution tide-gauges record sea-level variations at a 1-minute sampling rate using OTT radar-level sensors (RLS) with an accuracy of ±1 mm. A detailed list of the observations is provided in Table \ref{tab:tab02}.

\begin{table}[h]
\caption{MESSI observational network. AP indicates the presence of air pressure recorders and SL indicates sea-level tide-gauges.}\label{tab:tab02}
\begin{tabular}{@{}lccc@{}}
\toprule
Location & Coordinates & Observations \\
\midrule
Ancona (An)       & (13.506°E, 43.625°N) & AP \\
Ortona (Or)       & (14.415°E, 42.356°N) & AP \\
Vieste (Ve)       & (16.177°E, 41.888°N) & AP \\
Svetac (Sv)       & (15.757°E, 43.024°N) & AP \\
Vis (Vs)          & (16.192°E, 43.057°N) & AP \\
Vela Luka (VL)    & (16.703°E, 42.962°N) & AP, SL \\
Stari Grad (SG)   & (16.576°E, 43.181°N) & AP, SL\\
Vrboska (Vr)      & (16.671°E, 43.182°N) & AP \\
\bottomrule
\end{tabular}
\end{table}

The microbarographs are strategically deployed in regions known for the generation and amplification of meteotsunamis (marked by pink circles in Figure \ref{fig:fig01}), including Ancona (An), Ortona (Or), and Vieste (Ve) on the western Adriatic coast; Vis (Vs) and Svetac (Sv) on central Adriatic islands; and Vela Luka (VL), Stari Grad (SG), and Vrboska (Vr) on the eastern Adriatic coast. Tide-gauges are installed in Vela Luka and Stari Grad, both of which are harbors historically impacted by meteotsunami events. However, it is important to note that these tide-gauges are not located at the innermost points of the bays—typically where the strongest sea-level oscillations are observed—but approximately 2 km seaward. As a result, the recorded high-frequency sea-level signals are generally attenuated and may be two to three times smaller than those reported by eyewitnesses closer to the bay heads.

The 2014 and 2017 meteotsunami events occurred prior to the full deployment of the MESSI observational devices. Consequently, to complement this network, an additional dataset was sourced from Crometeo \citep{Crometeo2025}, a non-profit association of Croatian amateur meteorologists. This dataset provides air pressure records at 5-minute intervals from 39 locations across the Croatian coast and islands (indicated by white dots in Figure \ref{fig:fig01}). While coarser in resolution, this supplementary data offers valuable spatial coverage and historical insight into the synoptic and mesoscale conditions during earlier events. 

\subsection{Methods}\label{sec:sec2.5}

In this study, both modeled and observed air pressure and sea-level data—produced by the AdriSC modeling suite, the Meteo-HySEA model, and recorded by the MESSI and Crometeo stations—are processed using a 2-hour Kaiser–Bessel high-pass filter to isolate the high-frequency oscillations characteristic of meteotsunamis. A first-order validation is performed through a direct comparison between the high-pass-filtered time series of modeled and observed air pressure and sea level. This approach serves to evaluate the ability of the deterministic AdriSC and Meteo-HySEA models to capture meteotsunami signals at selected locations during the three selected events: 25–26 June 2014, 28 June–1 July 2017, and 11–19 May 2020. Additionally, in the AdriSC WRF 1.5-km model, the maximum pressure disturbances are characterized by the greatest temporal rate of change in high-pass-filtered air pressure, calculated as the pressure difference over a 5-minute interval. This method has been shown to effectively identify meteotsunamigenic disturbances \citep{Denamiel2019b}.

To characterize the dominant features of the atmospheric pressure disturbances recorded and modeled at the Crometeo stations, a zero-upcrossing method is applied to the 5-minute filtered air pressure time series. This method, commonly used in wave analysis, identifies individual wave-like events by detecting consecutive crossings of the pressure signal through the zero baseline in the upward direction. From these crossings, the wave height is defined as the difference between the maximum and minimum pressure values within each wave cycle, while the wave period corresponds to the time interval between two successive zero-upcrossings. This analysis allows for a systematic estimation of the pressure disturbance amplitude and temporal scale, which are critical parameters for assessing the potential of these disturbances to generate meteotsunamis. Applying this method to the Crometeo network and the WRF-ERAI and WRF-ERA5 results (resampled at 5-min temporal resolution) provides valuable additional comparison of the atmospheric disturbances (despite the lower temporal resolution compared to the MESSI stations) during the 2014 and 2017 meteotsunami event. The zero-upcrossing method is also applied to the 1-minute filtered sea-level series in order to compare the meteotsunami wave characteristics observed at the tide-gauges with both the AdriSC-ADCIRC and Meteo-HySEA model results. 

\section{Results}\label{sec:sec3}

\subsection{Atmospheric Forcing}\label{sec:sec3.1}

\begin{figure}
    \centering
    \includegraphics[width=0.75\linewidth]{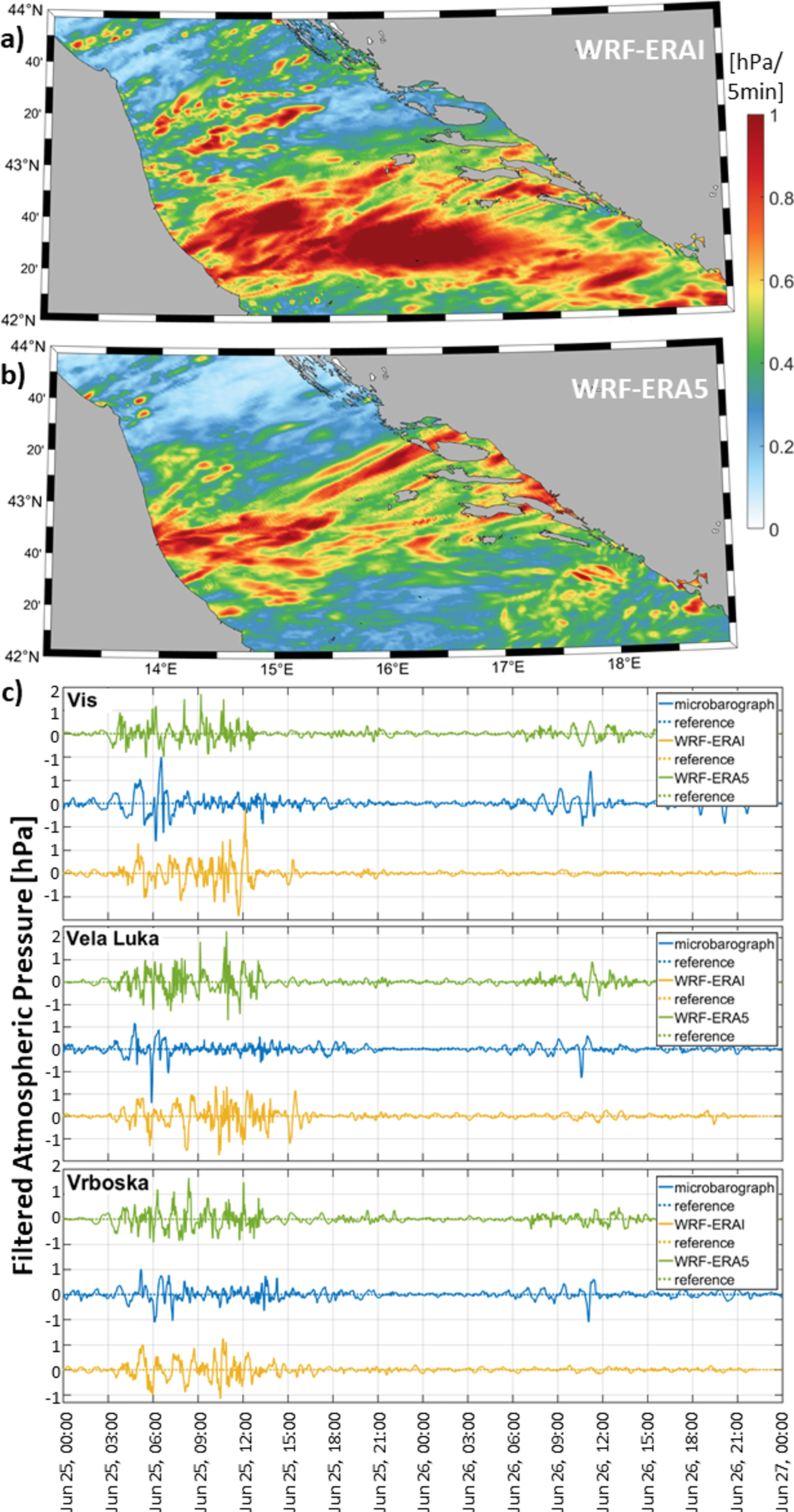}
    \caption{Meteotsunami events of June 25 and 26 2014. Spatial distribution of the maximum pressure disturbances for (a) the WRF-ERAI and (b) WRF-ERA5 simulations. In panel (c), observed and modeled filtered atmospheric pressure at the Vis, Vela Luka and Vrboska MESSI microbarograph locations.}
    \label{fig:fig04}
\end{figure}

\begin{figure}
    \centering
    \includegraphics[width=0.75\linewidth]{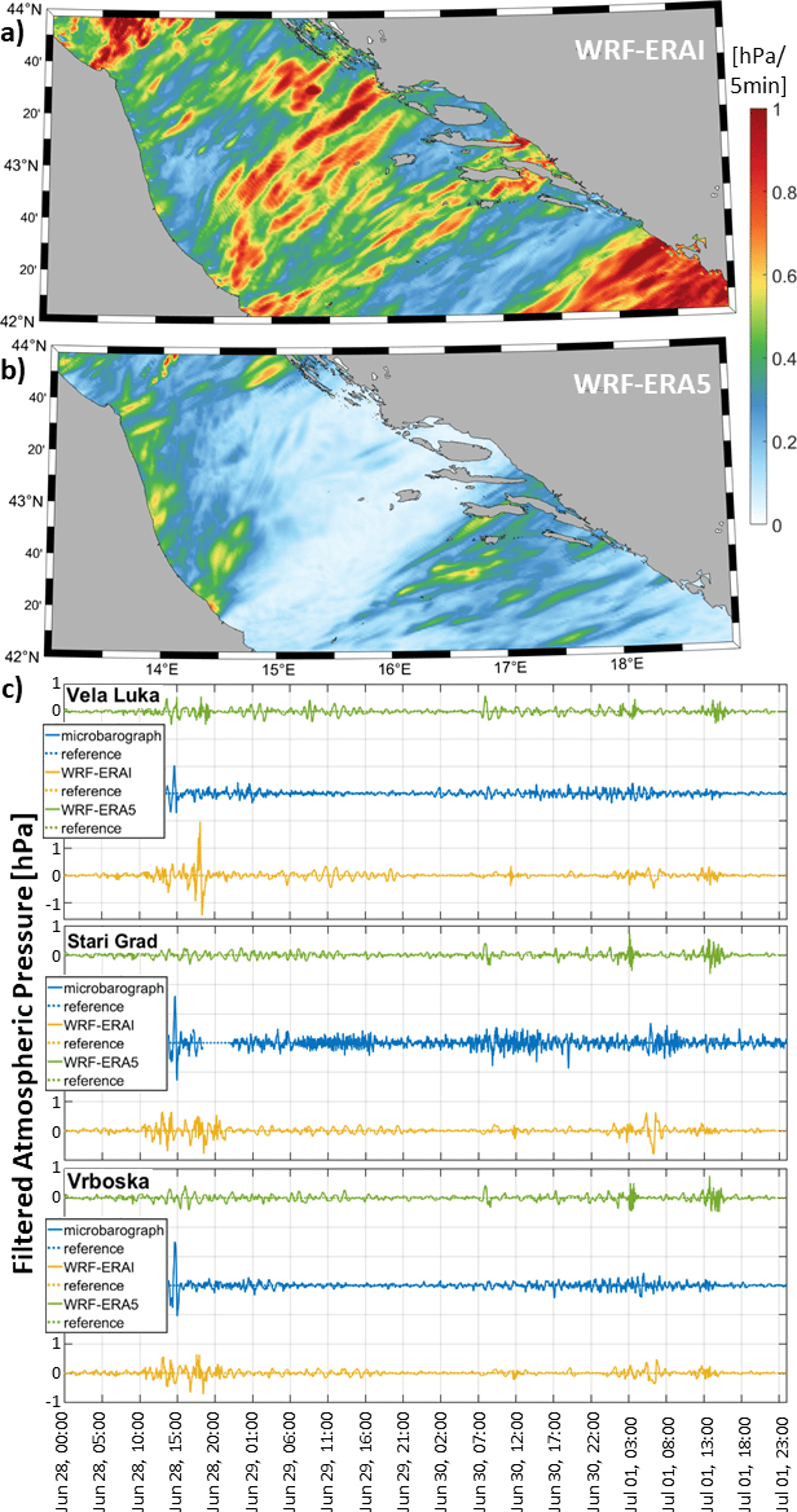}
    \caption{Meteotsunami events from June 28 to July 02 2017. Spatial distribution of the maximum pressure disturbances for (a) the WRF-ERAI and (b) WRF-ERA5 simulations. Panel (c) compares observed and modeled filtered atmospheric pressure at the Vela Luka, Stari Grad and Vrboska MESSI microbarograph locations.}
    \label{fig:fig05}
\end{figure}

To assess the ability of the AdriSC model to reproduce atmospheric disturbances responsible for meteotsunami events in the Adriatic Sea, we first analyze the spatial distribution of maximum atmospheric pressure disturbances and the temporal evolution of high-pass filtered pressure signals at several microbarograph stations (Vis, Vela Luka, Stari Grad, and Vrboska). Specifically, we compare available observations with outputs from WRF simulations at 1 km resolution, forced by both ERA-Interim (ERAI) and ERA5 reanalyses, for the events of June 25–26, 2014 (Fig. \ref{fig:fig04}), June 28–July 2, 2017 (Fig. \ref{fig:fig05}), and May 11–19, 2020 (Fig. \ref{fig:fig06}). Additional comparisons of observed and modeled pressure disturbances are provided in Appendix A (Figs. \ref{fig:figA1}–\ref{fig:figA3}).

For the June 2014 event (Fig. \ref{fig:fig04}a,b), the WRF-ERA5 simulation produces weaker and more spatially localized pressure disturbances, with most maxima below 0.5 hPa/5 min. In contrast, the WRF-ERAI simulation generates more intense and widespread disturbances, exceeding 0.8 hPa/5 min across much of the Adriatic basin. The time series comparison (Fig. \ref{fig:fig04}c) shows that WRF-ERA5 captures several of the pressure fluctuations observed at Vis and Vela Luka, albeit with lower amplitude and broader temporal structure than the observations. Conversely, WRF-ERAI tends to overestimate the frequency and intensity of disturbances, particularly at Vis, where it simulates numerous spurious events absent from observations—suggesting a tendency for over-amplification of mesoscale pressure features in the ERAI-driven simulation.

\begin{figure}
    \centering
    \includegraphics[width=0.75\linewidth]{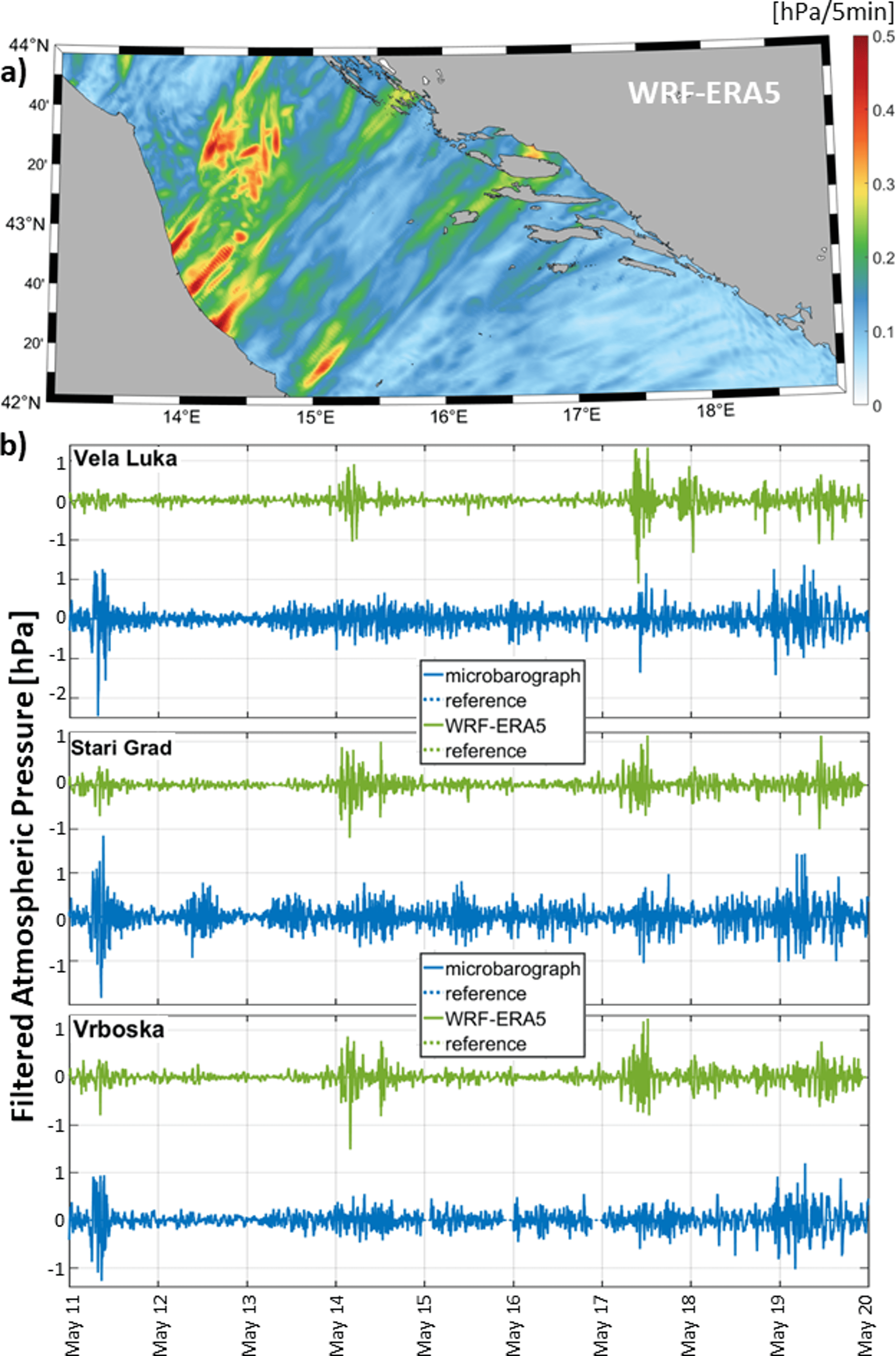}
    \caption{Meteotsunami events of May 11-19 2020. (a) Spatial distribution of the maximum pressure disturbances for the WRF-ERA5 simulation. (b) Observed and modeled filtered atmospheric pressure at the Vela Luka, Stari Grad and Vrboska MESSI microbarograph locations.}
    \label{fig:fig06}
\end{figure}

For the June–July 2017 event (Fig. \ref{fig:fig05}a,b), the WRF-ERAI simulation once again produces more widespread and intense pressure disturbances, with anomalies reaching up to 0.8 hPa/5 min. In contrast, the WRF-ERA5 simulation generates weaker and more localized pressure fluctuations, generally below 0.5 hPa/5 min. The comparison of the pressure time series (Fig. \ref{fig:fig05}c) indicates that WRF-ERA5 performs significantly better than WRF-ERAI in capturing the strongest observed atmospheric disturbance on June 28, which peaked near 2.5 hPa. However, neither WRF-ERA5 nor WRF-ERAI succeed in reproducing the persistent high-amplitude fluctuations (ranging from 0.5 to 0.75 hPa) recorded by the Stari Grad microbarograph between June 29 and July 1, suggesting limitations in both simulations during the final stages of the event.

For the May 2020 event (Fig. \ref{fig:fig06}a), the WRF-ERA5 simulation shows maximum pressure disturbances of up to approximately 0.5 hPa/5 min, primarily concentrated between the Italian coastline and the central Dalmatian coast. These features are more localized and of lower amplitude compared to those modeled with WRF-ERAI in the 2014 and 2017 events. The corresponding time series (Fig. \ref{fig:fig06}b) reveal that WRF-ERA5 captures the timing and magnitude of several key pressure fluctuations observed at Vela Luka, Stari Grad, and Vrboska between May 14 and May 17. However, the model fails to reproduce the persistent oscillations in the range of 0.5 to 1 hPa observed during this period, and more critically, completely misses the strongest disturbance—a sharp pressure jump reaching up to 2.5 hPa—recorded on May 11. This highlights a significant limitation in the model’s ability to capture the full intensity of atmospheric disturbances associated with this particular event.

In summary, the comparison between WRF-1km simulations forced with ERAI and ERA5 reanalyses highlights notable differences in their ability to reproduce the atmospheric disturbances responsible for meteotsunami generation in the Adriatic Sea. While WRF-ERAI systematically generates more intense and widespread pressure anomalies—sometimes exceeding observed amplitudes—it likely over-amplifies mesoscale features and produces spurious fluctuations. In contrast, WRF-ERA5 generally yields more localized and lower-amplitude disturbances, which align more closely with observations in terms of timing and structure. However, ERA5-driven simulations sometimes fail to capture the strongest pressure anomalies or their precise spatial positioning. Due to the scarcity of observational data over the central Adriatic, between the Italian and Croatian coasts where key atmospheric features may develop, the evaluation of the results produced by WRF-ERAI and WRF-ERA5 remains largely incomplete and can not provide enough information to properly assess the skills of the models. Consequently, given that ERAI has been discontinued and no longer provides a viable option for future research, the following analyses focus mostly on simulations forced with ERA5, which, despite its limitations, offers the best available representation of atmospheric conditions over the Adriatic region.

\subsection{Performance of the Meteo-HySEA model}\label{sec:sec3.2}

\begin{figure}
    \centering
    \includegraphics[width=0.75\linewidth]{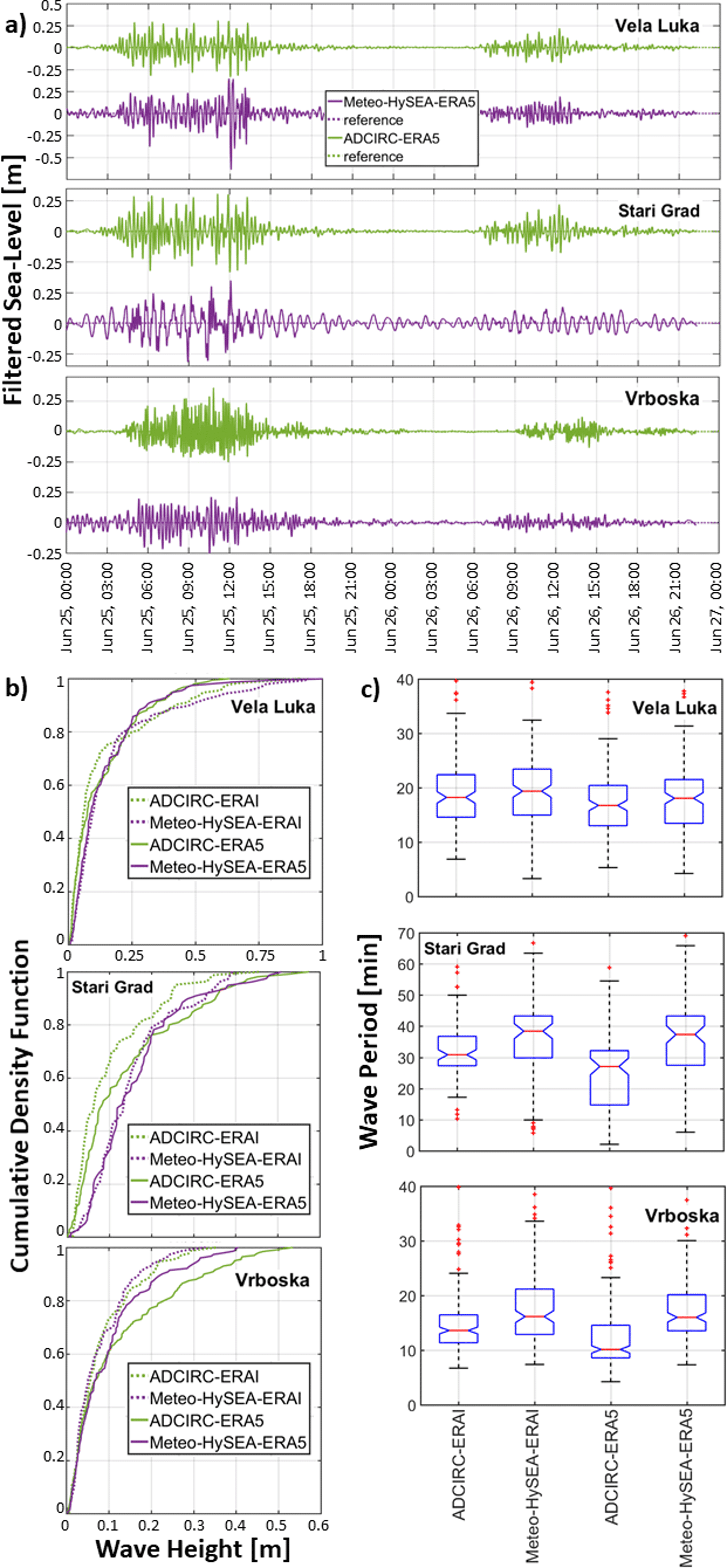}
    \caption{Meteotsunami events of June 25 and 26 2014 at Vela Luka, Stari Grad and Vrboska locations. (a) Time series of modeled filtered sea-levels. (b) Cumulative Density Function (CDF) of the wave heights and (c) box plots of the wave periods extracted from the time series.}
    \label{fig:fig07}
\end{figure}

\begin{figure}
    \centering
    \includegraphics[width=0.75\linewidth]{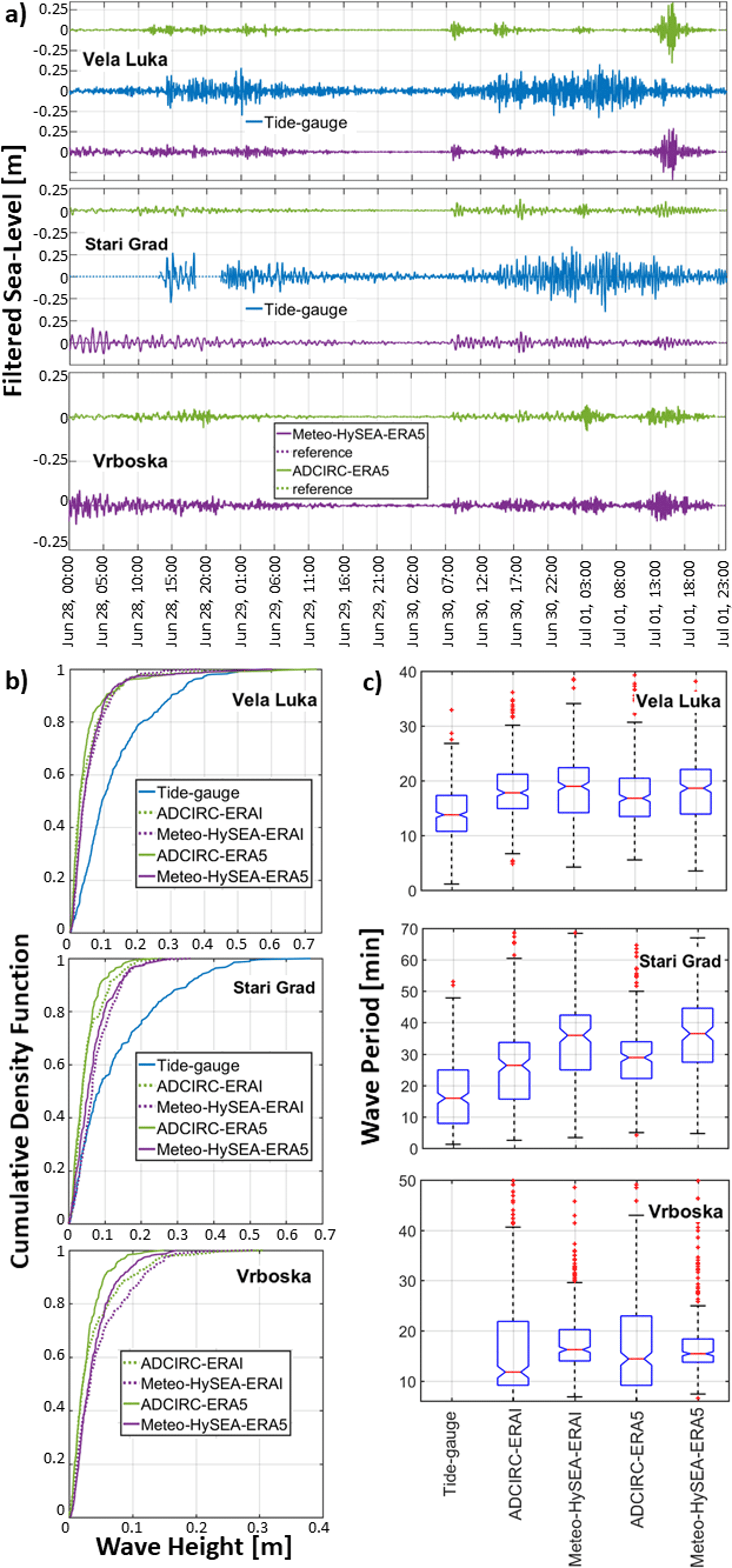}
    \caption{Meteotsunami events of June 28 to July 1 2017 at Vela Luka, Stari Grad and Vrboska locations. (a) Time series of modeled filtered sea-levels. (b) Cumulative Density Function (CDF) of the wave heights and (c) box plots of the wave periods extracted from the time series.}
    \label{fig:fig08}
\end{figure}

\begin{figure}
    \centering
    \includegraphics[width=0.75\linewidth]{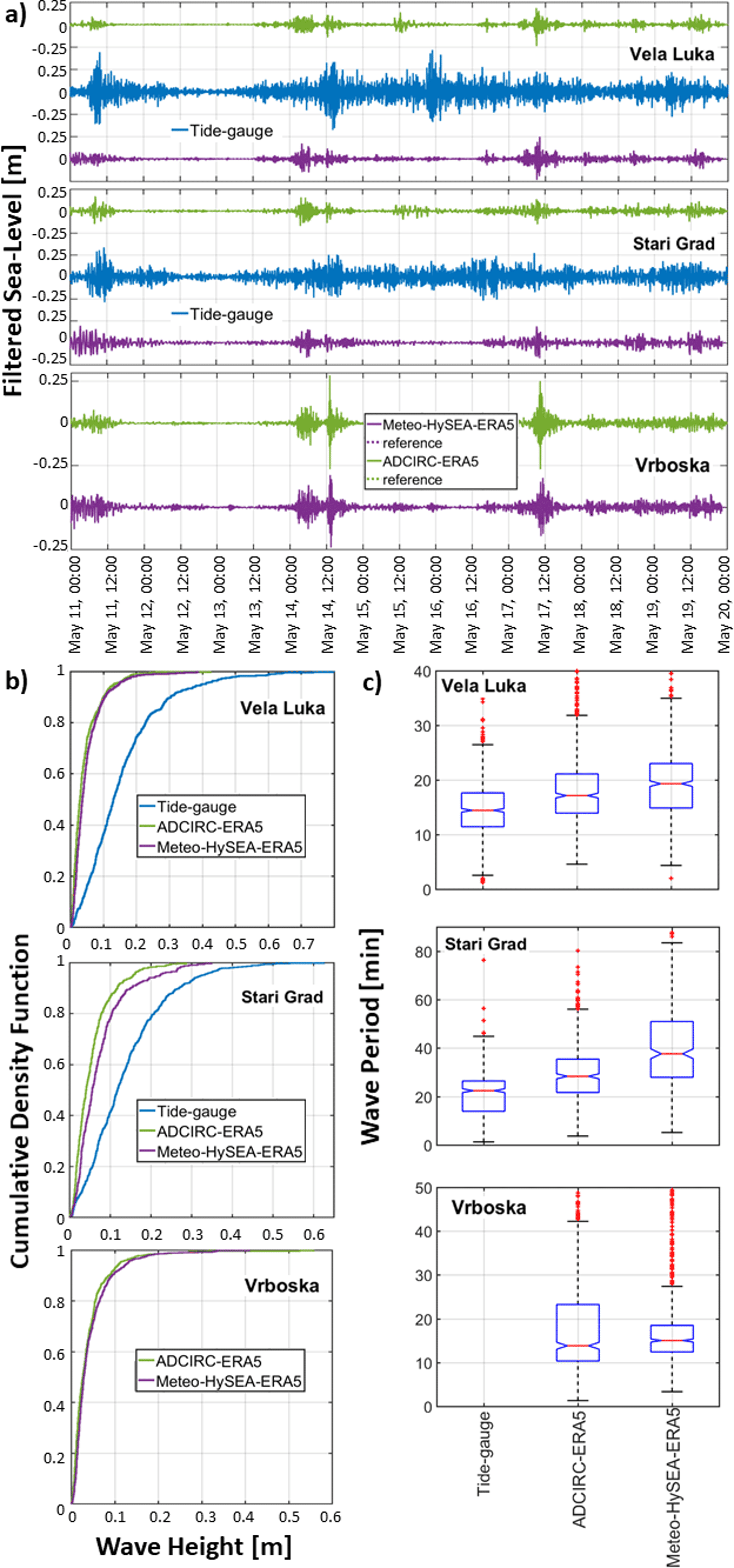}
    \caption{Meteotsunami events of May 11-19 2020 at Vela Luka, Stari Grad and Vrboska locations. (a) Time series of modeled filtered sea-levels. (b) Cumulative Density Function (CDF) of the wave heights and (c) box plots of the wave periods extracted from the time series.}
    \label{fig:fig09}
\end{figure}

In this section, we present a comparative analysis of the performance of the newly developed Meteo-HySEA model against the widely used AdriSC-ADCIRC, with both models evaluated against available tide gauge observations at Vela Luka and Stari Grad for the three studied meteotsunami events: June 25–26, 2014, June 28–July 1, 2017, and May 11–19, 2020. The evaluation includes filtered sea-level time series and distributions of wave height and wave periods (Figs. \ref{fig:fig07}–\ref{fig:fig09}). For completeness, a comparison of the filtered sea-level time series obtained using the WRF-ERAI forcing is also provided in Appendix B (Figs. \ref{fig:figB4}, \ref{fig:figB5}).

The June 2014 events (Fig. \ref{fig:fig07}a, WRF-ERA5 forcing only) are characterized by sharp oscillations within the Vela Luka, Stari Grad, and Vrboska harbors. On June 25, compared to AdriSC-ADCIRC, Meteo-HySEA reproduces stronger peak amplitudes in Vela Luka (approximately 1 m vs. 0.6 m) but weaker amplitudes in Stari Grad (0.5 m vs. 0.6 m) and Vrboska (0.4 m vs. 0.5 m). For the June 26 event, Meteo-HySEA slightly underestimates the oscillations in Vela Luka (0.3 m vs. 0.4 m) and Vrboska (0.15 m vs. 0.2 m), while in Stari Grad both models produce similar oscillations of about 0.19 m. The cumulative density functions (CDFs) of wave heights (Fig. \ref{fig:fig07}b) confirm these results, highlighting that Meteo-HySEA, whether forced by WRF-ERAI or WRF-ERA5, consistently produces stronger oscillations than AdriSC-ADCIRC in Vela Luka but weaker oscillations in Vrboska. In Stari Grad, Meteo-HySEA tends to generate a higher number of oscillations below 0.2 m compared to AdriSC-ADCIRC. Finally, for both ERAI and ERA5 forcing, Meteo-HySEA consistently overestimates the periods of meteotsunami oscillations relative to AdriSC-ADCIRC (Fig. \ref{fig:fig07}c). For instance, using WRF-ERA5 forcing, it produces median periods of 18 min (vs. 17 min) in Vela Luka, 37 min (vs. 27 min) in Stari Grad, and 16 min (vs. 10 min) in Vrboska.

For the June–July 2017 events (Fig. \ref{fig:fig08}a, WRF-ERA5 forcing only), both Meteo-HySEA and AdriSC-ADCIRC fail to reproduce the long-lasting oscillations above 0.15 m and up to 0.7 m observed at Vela Luka and Stari Grad during June 28, June 30, and July 1. At Vrboska, oscillations produced by both models remain mostly below 0.15 m. The highest modeled oscillations occur at the end of July 1, reaching approximately 0.6 m (vs. 0.7 m observed) in Vela Luka, 0.3 m (vs. 0.25 m) in Stari Grad, and 0.2 m (vs. 0.15 m) in Vrboska. The CDFs (Fig. \ref{fig:fig08}b) confirm these results: more than 20\% of the observed oscillations exceed 0.2 m at Vela Luka and Stari Grad, whereas fewer than 5\% of wave heights modeled by Meteo-HySEA or AdriSC-ADCIRC (forced by either WRF-ERAI or WRF-ERA5) exceed this threshold. Similar to the June 2014 events, Meteo-HySEA overestimates the median wave period of the meteotsunamis compared to AdriSC-ADCIRC (Fig. \ref{fig:fig08}c): 19 min (vs. 17 min) in Vela Luka, 36 min (vs. 29 min) in Stari Grad, and 15 min (vs. 14 min) in Vrboska. However, observed median wave periods are 14 min in Vela Luka and 16 min in Stari Grad, suggesting that the harbor geomorphology is likely not well captured by either model, particularly in Stari Grad.

For the May 2020 events (Fig. \ref{fig:fig09}a, WRF-ERA5 forcing only), similar to the June–July 2017 events, neither Meteo-HySEA nor AdriSC-ADCIRC reproduce the sustained oscillations above 0.5 m observed at Vela Luka and Stari Grad. The maximum sea-level oscillations modeled by Meteo-HySEA and AdriSC-ADCIRC reach approximately 0.4 m in Vela Luka and 0.35 m and 0.3 m, respectively, in Stari Grad, whereas observations reach 0.8 m in Vela Luka and 0.6 m in Stari Grad. Furthermore, the CDFs (Fig. \ref{fig:fig09}b) indicate that 20\% of the observed wave heights in Vela Luka and Stari Grad exceed 0.2 m, while less than 1\% and 10\% of the Meteo-HySEA wave heights exceed this threshold in Vela Luka and Stari Grad, respectively. At Stari Grad, AdriSC-ADCIRC performs even worse than Meteo-HySEA, with less than 5\% of wave heights exceeding the 0.2 m threshold. As for the other events, Meteo-HySEA overestimates the meteotsunami periods compared to AdriSC-ADCIRC (Fig. \ref{fig:fig09}c): 19 min (vs. 17 min) in Vela Luka, 37 min (vs. 28 min) in Stari Grad, and 15 min (vs. 14 min) in Vrboska. However, observed median wave periods are 14 min in Vela Luka and 22 min in Stari Grad.

In summary, the analyses of Meteo-HySEA sea-levels overall highlight the model skills in reproducing the temporal evolution and amplitude of meteotsunami-induced oscillations produced by the AdriSC-ADCIRC model. However, the comparison with tide-gauge observations for the 2017 and 2020 events reveals that both Meteo-HySEA and AdriSC-ADCIRC models significantly underestimate the observed amplitudes at Vela Luka and Stari Grad, with discrepancies reaching more than 50\%. This underestimation is consistent with the lower atmospheric pressure disturbance amplitudes produced by the WRF-ERA5 and WRF-ERAI simulations for these events (Figs. \ref{fig:fig05}, \ref{fig:fig06}). Furthermore, across all events and locations, Meteo-HySEA generally produces longer sustained sea-level oscillations than AdriSC-ADCIRC, which may indicate either a more effective response to the available atmospheric forcing or suggest that general circulation, tidal and wind forcing—included in AdriSC-ADCIRC but not in the Meteo-HySEA simulations—play a non-negligible role in the propagation and amplification of small meteotsunami waves. Concerning the wave periods derived from zero-crossing analysis, they indicate that Meteo-HySEA generally simulates longer wave periods than AdriSC-ADCIRC, with both models systematically overestimating the observed values. In particular, Meteo-HySEA shows a greater deviation at Stari Grad, where its median wave periods exceed observed values by up to 50\%. This suggests that basin-specific resonance characteristics and the timing of meteotsunami wave propagation are either influenced by the inclusion of general circulation, tidal and wind forcing in AdriSC-ADCIRC, or are less accurately represented in Meteo-HySEA compared to AdriSC-ADCIRC.

\begin{figure}
    \centering
    \includegraphics[width=0.9\linewidth]{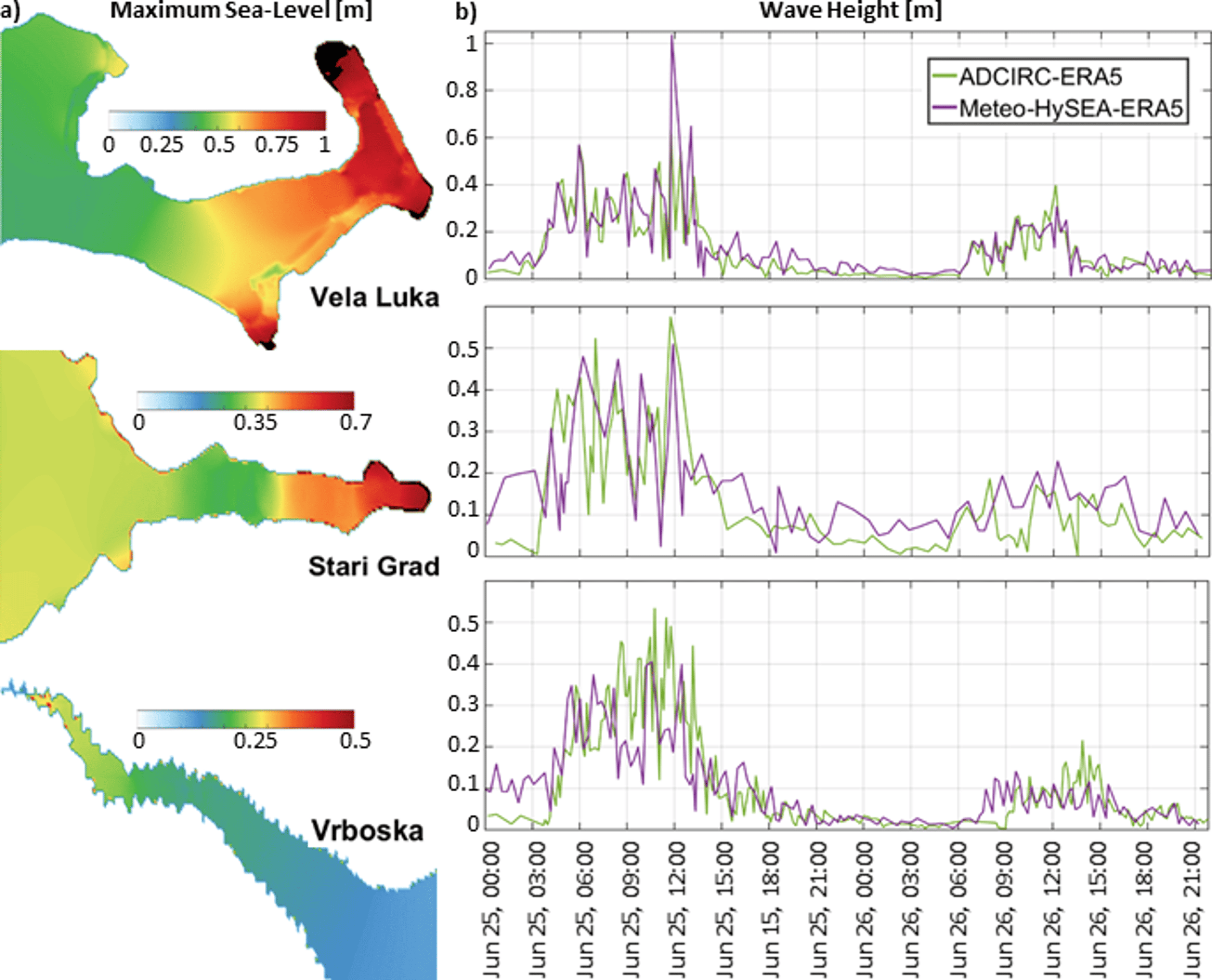}
    \caption{Meteotsunami events of June 25 and 26 2014. (a) Maximum sea-levels and (b) time series of wave height extracted from the 1-min filtered sea-level results of Meteo-HySEA forced by WRF-ERA5 for the three 7 m grids of Vela Luka, Stari Grad and Vrboska. The time series of wave height are also compared with the results extracted from AdriSC-ADCIRC forced by WRF-ERA5.}
    \label{fig:fig10}
\end{figure}

\section{Discussion}\label{sec:sec4}

Beyond the direct comparison with observations presented in the results, a deeper evaluation of Meteo-HySEA’s skill is required. In this section, we therefore expand the analysis by contrasting Meteo-HySEA with AdriSC-ADCIRC, considering both the real events introduced earlier and a set of three synthetic pressure disturbances designed to generate the strongest plausible meteotsunamis in the studied bays. While the real events provide an opportunity to highlight differences in model behavior under observed conditions, the synthetic experiments allow for a controlled assessment of performance under idealized but physically meaningful forcing. Taken together, these complementary perspectives provide a more comprehensive evaluation of Meteo-HySEA’s ability to reproduce meteotsunamis in the Adriatic Sea.

\subsection{Real Meteotsunami Events}\label{sec:sec4.1}

\begin{figure}
    \centering
    \includegraphics[width=0.9\linewidth]{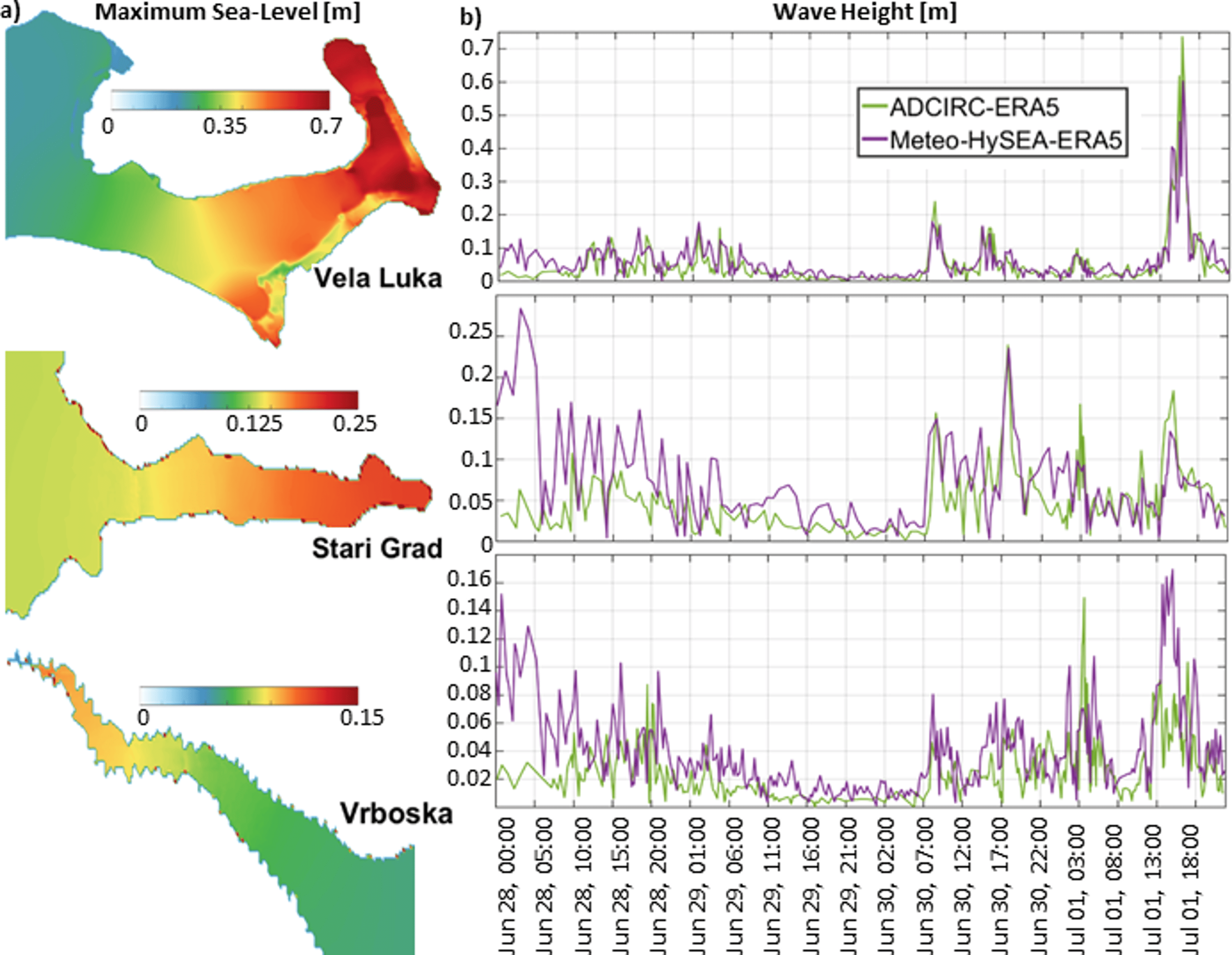}
    \caption{Meteotsunami events of June 28 to July 1 2017. (a) Maximum sea-levels and (b) time series of wave height extracted from the 1-min filtered sea-level results and Meteo-HySEA forced by WRF-ERA5 for the three 7 m grids of Vela Luka, Stari Grad and Vrboska. The time series of wave height are also compared with the results extracted from AdriSC-ADCIRC forced by WRF-ERA5.}
    \label{fig:fig11}
\end{figure}

\begin{figure}
    \centering
    \includegraphics[width=0.9\linewidth]{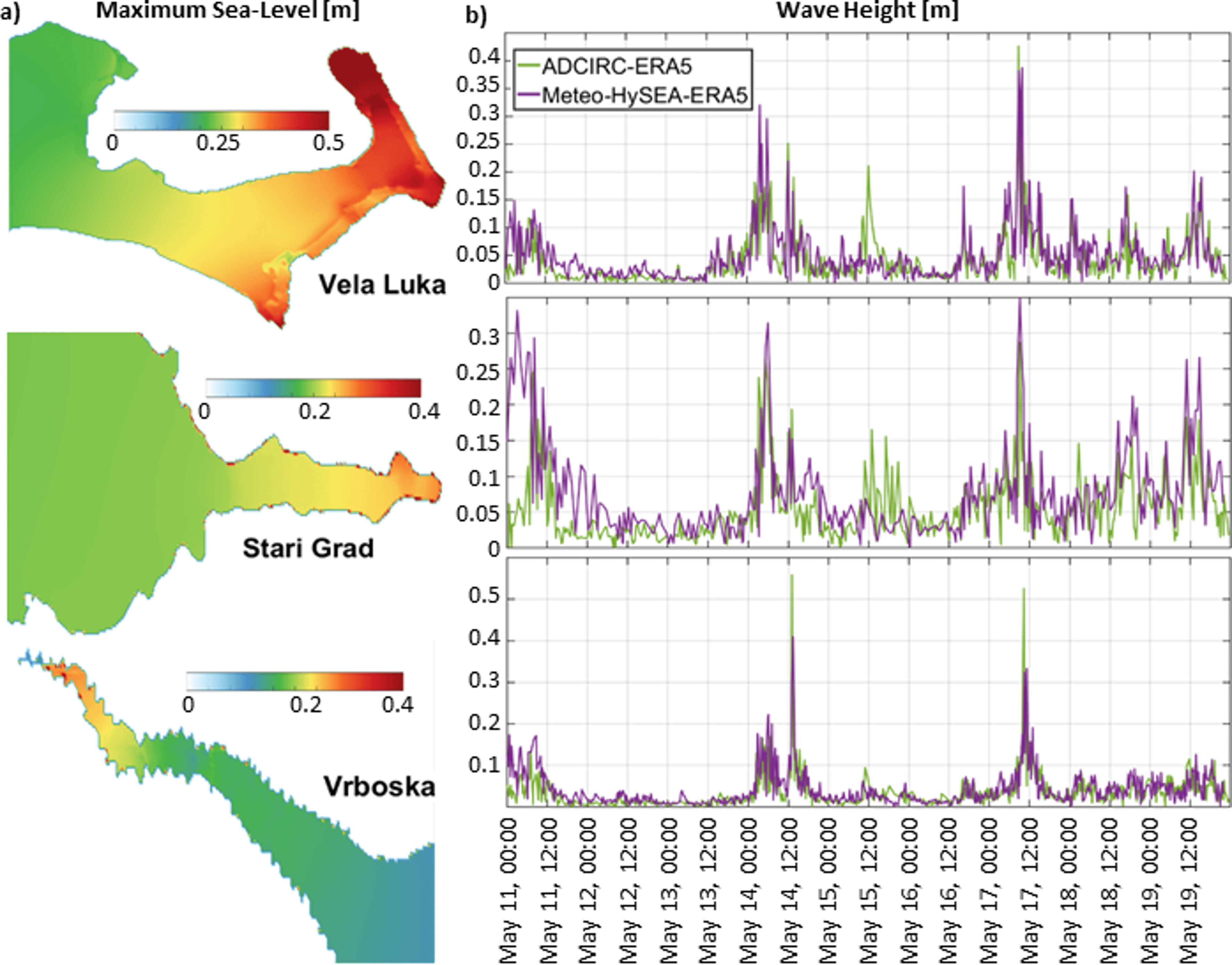}
    \caption{Meteotsunami events of May 11-19 2020. (a) Maximum sea-levels and (b) time series of wave height extracted from the 1-min filtered sea-level results and Meteo-HySEA forced by WRF-ERA5 for the three 7 m grids of Vela Luka, Stari Grad and Vrboska. The time series of wave height are also compared with the results extracted from AdriSC-ADCIRC forced by WRF-ERA5.}
    \label{fig:fig12}
\end{figure}

The ability of Meteo-HySEA to reproduce both the temporal evolution and spatial distribution of sea-level oscillations in the eastern Adriatic bays of Vela Luka, Stari Grad, and Vrboska has been demonstrated based on three real meteotsunami events. While the overall agreement with AdriSC-ADCIRC is encouraging, systematic differences arise, particularly regarding the amplitude and period of the simulated oscillations. To further discuss and assess these differences for the WRF-ERA5 forcing, we examine Meteo-HySEA performance across the three bays by analyzing the spatial patterns of maximum sea levels as well as the time series of wave heights in comparison with AdriSC-ADCIRC, as illustrated in Figures \ref{fig:fig10}–\ref{fig:fig12}. The comparison of the three events underscores both the strengths and limitations of Meteo-HySEA relative to AdriSC-ADCIRC. 

For the intense June 25–26, 2014 events (Fig. \ref{fig:fig10}), both models reproduce the same timing of the strongest disturbances, confirming their ability to capture the underlying resonance processes in Vela Luka and Stari Grad. However, whereas AdriSC-ADCIRC produced slightly more diffusive signals, Meteo-HySEA tends to yield higher and more variable wave heights, particularly in the cases of Vela Luka and Stari Grad, with a clear overestimation at the onset of the event in Stari Grad and Vrboska. This early amplification, not supported by AdriSC-ADCIRC, may be due to the absence of wind and tidal updates in the Meteo-HySEA simulations, whereas AdriSC-ADCIRC incorporates these boundary conditions on an hourly basis. The time series shown in Figures \ref{fig:fig10}b–\ref{fig:fig12}b start after twelve hours of simulation—the spin-up period. By this time, AdriSC-ADCIRC has already assimilated twelve tidal and wind updates, while Meteo-HySEA still relies on the sea state prescribed twelve hours earlier.

A similar pattern emerged during the weaker and more prolonged June 28 – July 1, 2017 event (Fig. \ref{fig:fig11}). Both models reproduced the modest amplification observed in Stari Grad and Vrboska, as well as the stronger oscillations in Vela Luka. However, Meteo-HySEA overestimated the largest peaks, particularly during the later stages of the event at Vrboska. Moreover, the timing of the major peaks in the 2014 and 2017 events was not well captured at this location. This discrepancy is likely related to the nearshore resolution, as the narrow bay of Vrboska is not adequately resolved by the grid employed in the Meteo-HySEA simulations. As in 2014, the initial response in the narrower bays was unrealistically strong, suggesting that this may be a systematic issue rather than event-specific behavior.

The May 11–19, 2020 event (Fig. \ref{fig:fig12}) highlights another dimension of model performance. This long-lasting event, characterized by moderate but recurrent bursts of activity, was more uniformly distributed across the three bays. Both models reproduced the persistence and general amplitude of the oscillations; however, Meteo-HySEA systematically generated more energetic signals, particularly at Stari Grad. In Vrboska, the largest peaks simulated with Meteo-HySEA underestimate those obtained with AdriSC-ADCIRC, once again pointing to possible limitations associated with the insufficiently resolved bathymetry in this area. In Stari Grad, the exaggerated initial oscillations were even more pronounced than in previous events, further  highlighting the limitations in how Meteo-HySEA handles the onset of disturbances in narrower, semi-enclosed basins.

Taken together, these comparisons indicate that Meteo-HySEA tends to reproduce peak oscillations more forcefully than AdriSC-ADCIRC, thereby probably better capturing the potential extremes of meteotsunami events. However, its higher short-term variability and tendency to overestimate the early response in certain bays remain recurring shortcomings. The exaggerated initial oscillations suggest that the inundation and wetting–drying schemes may require a longer spin-up period to stabilize before the main oscillations develop. Alternatively, they may reflect sensitivity to: (1) general circulation, tidal and wind forcing, (2) small-scale atmospheric forcing fluctuations or (3) numerical reflections or friction parametrization within the bays. Recognizing and addressing these issues would improve Meteo-HySEA’s reliability and robustness, especially for operational forecasting applications where accurate early warning is critical.

The general overestimation of wave periods observed in Figures \ref{fig:fig07}c–\ref{fig:fig09}c can be attributed to several factors. First, it is largely influenced by the local geomorphology. In Meteo-HySEA, the wet–dry technique used to simulate inundation of dry cells causes the effective geomorphology to evolve over time, which may lead to inaccuracies in the estimation of the resonance period. Second, the coastal bathymetric coverage is rather limited, which further affects the wet–dry technique and, consequently, the calculation of the wave period.

\subsection{Synthetic Meteotsunami Events}\label{sec:sec4.2}

\begin{figure}
    \centering
    \includegraphics[width=1.0\linewidth]{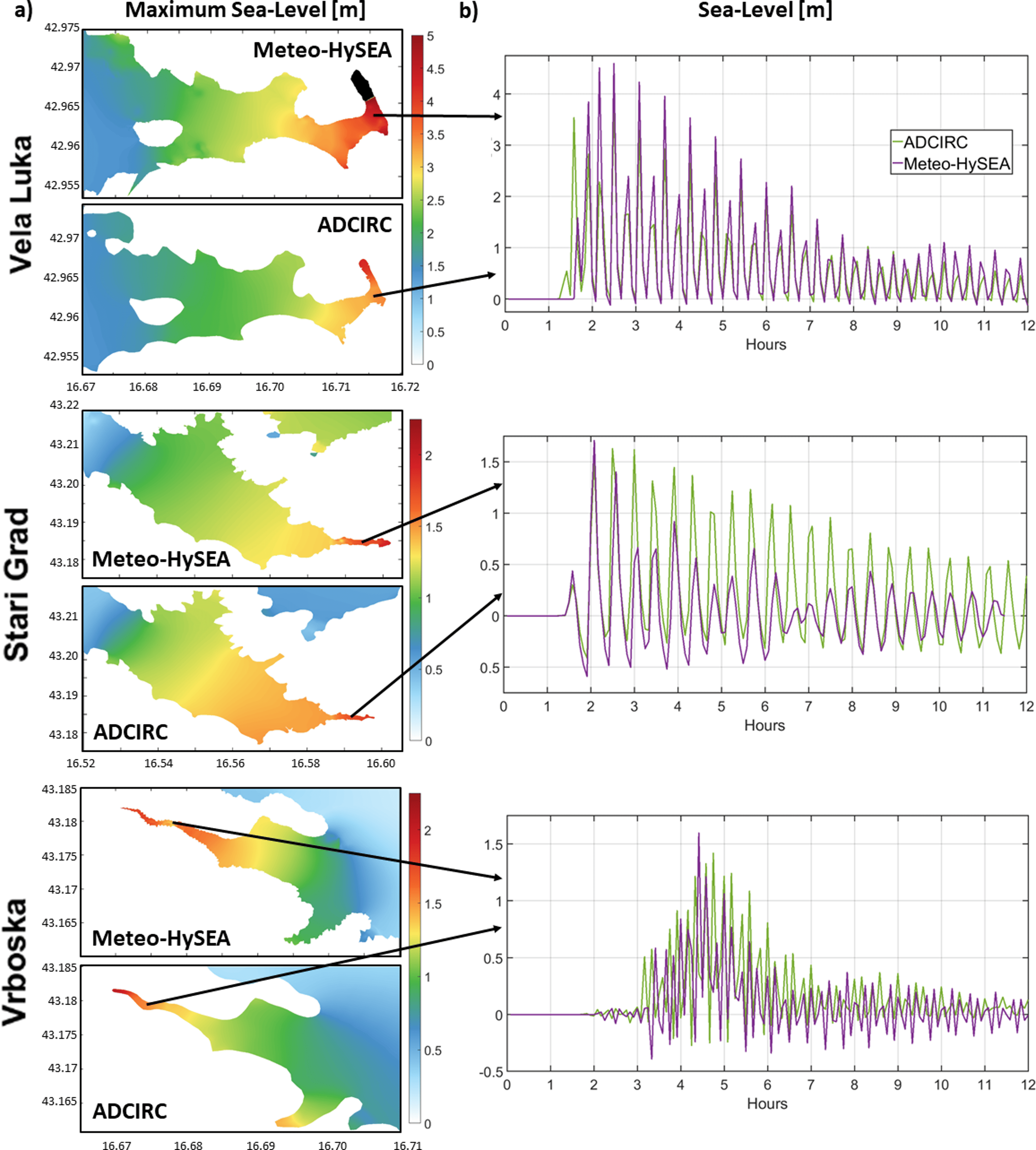}
    \caption{(a) Spatial distributions of maximum sea-levels and (b) time series of sea-levels during the three synthetic meteotsunami events with the highest impacts on the harbors of Vela Luka, Stari Grad and Vrboska.}
    \label{fig:fig13}
\end{figure}

\begin{figure}
    \centering
    \includegraphics[width=0.6\linewidth]{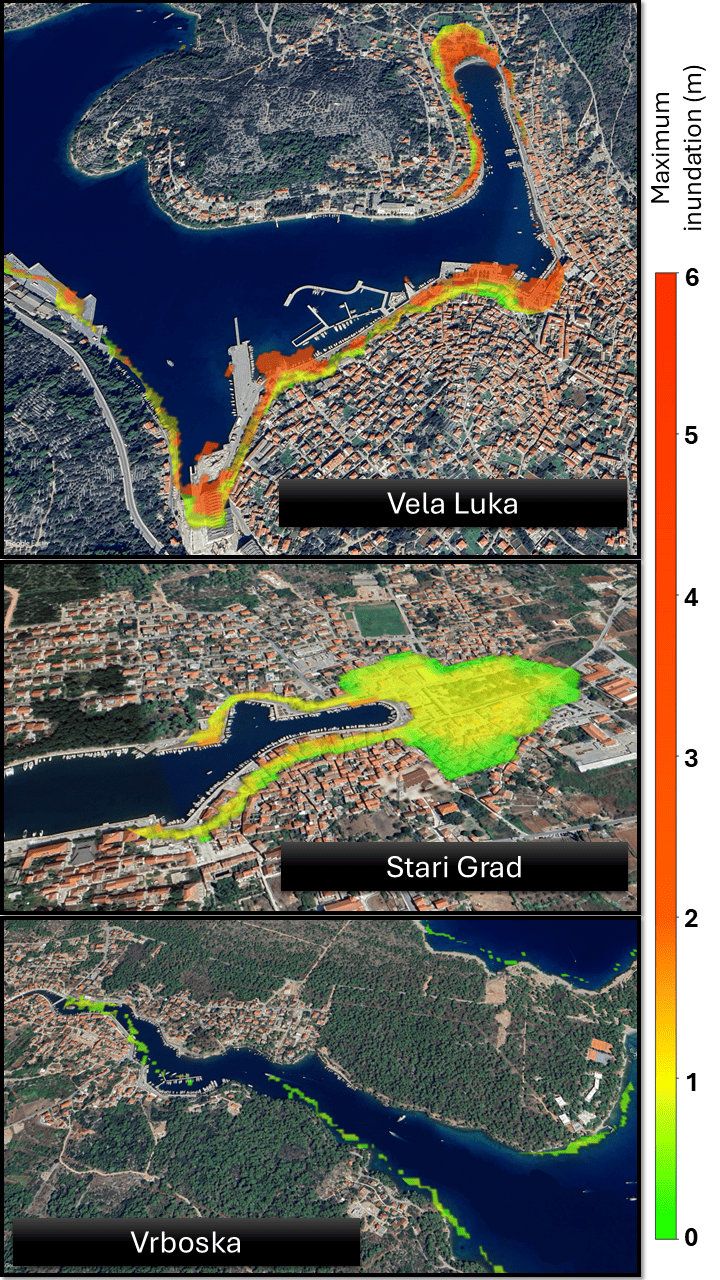}
    \caption{Spatial distributions of maximum inundation during the three synthetic meteotsunami events with the highest impacts on the harbors of Vela Luka, Stari Grad and Vrboska.}
    \label{fig:fig14}
\end{figure}

 Given the challenges of reproducing real-world atmospheric disturbances and other external forcing (e.g., general circulation, tides and wind) responsible for meteotsunamis in the Adriatic Sea—as seen in the previous paragraphs—we further evaluate the Meteo‑HySEA performances using synthetic pressure forcing based on the analytical expression developed for the AdriSC meteotsunami surrogate model \citep{Denamiel2018, Denamiel2019b, Denamiel2020}. That synthetic framework was rigorously established through uncertainty quantification via polynomial chaos expansions applied to over 10,600 simulated pressure disturbances, enabling a robust mapping between atmospheric pressure parameters (point of origin, direction of propagation, amplitude, speed, period and width of the atmospheric disturbance) and extreme sea-level responses. Stochastic surrogate modeling has also demonstrated remarkable computational efficiency—executing more than three orders of magnitude faster than deterministic ocean models—while improving the reliability of hazard assessments for Adriatic meteotsunamis \citep{Denamiel2019b, Tojcic2021}.

For this study, three different synthetic pressure forcing cases, that produced the highest sea-level impacts in the harbors of Vela Luka, Stari Grad, and Vrboska are selected. The physical parameters of these scenarios are summarized in Table \ref{tab:tab03}. It should be noted that no other forcing than the pressure disturbances is used in both Meteo-HySEA and AdriSC-ADCIRC simulations.

\begin{table}[h]
\caption{Parameters of the atmospheric disturbances used to produce the synthetic meteotsunami events.}\label{tab:tab03}
\begin{tabular}{@{}lllllll@{}}
\toprule
         & Origin  & Direction & Amplitude & Speed  & Period  & Width\\
Location & (°N,°E) & (degrees) & (hPa)     & (m/s)  & (min.)  & (km)\\
\midrule
Vela Luka    & 41.30, 15.98 & 73 & 3.6 & 37.2 & 17.5 & 90.0\\
Stari Grad   & 41.52, 15.43 & 73 & 3.6 & 37.2 & 27.2 & 136.5\\
Vrboska      & 42.45, 14.10 & 15 & 3.9 & 27.5 & 12.0 & 136.5\\
\botrule
\end{tabular}
\end{table}

The comparison between Meteo-HySEA and AdriSC-ADCIRC for these three extreme synthetic forcing (Fig. \ref{fig:fig13}) highlights both consistent patterns and systematic differences in simulating meteotsunami dynamics across the three selected harbors. For maximum sea-level maps, Meteo-HySEA tends to produce higher amplitudes than AdriSC-ADCIRC, particularly in the innermost parts of the basins, except within the Stari Grad bay. This difference is most pronounced in Vela Luka, where Meteo-HySEA predicts extreme elevations exceeding 5 m, whereas AdriSC-ADCIRC peaks closer to 4 m (Fig. \ref{fig:fig13}a). These spatial differences suggest that Meteo-HySEA may enhance resonance effects more strongly than ADCIRC in the Vela Luka and Vrboska bays. The temporal series confirm these tendencies. In Vela Luka, both models capture the onset time and overall oscillatory behavior, but Meteo-HySEA yields higher peak amplitudes and a slower decay of oscillations compared to AdriSC-ADCIRC, which damps the signal more rapidly. A different behavior is observed in Stari Grad, where Meteo-HySEA reproduces sustained seiche-like oscillations of lower amplitude, while AdriSC-ADCIRC shows lower energy dissipation. In Vrboska, the two models are in closer agreement for the initial peaks, yet differences in the persistence of oscillations remain. These results indicate that, although both models successfully reproduce the timing and frequency of meteotsunami resonances, their ability to retain energy differs. This suggests that the the better treatment of the inundation implemented in Meteo-HySEA may also influence the results within the bays. Furthermore, differences in the representation of nearshore geomorphology between the two models likely affect the oscillations recorded at the points of interest. While AdriSC-ADCIRC relies on an unstructured mesh with high-resolution and accurate data, Meteo-HySEA requires multiple reinterpolations of these values onto the 7\,m grids, leading to less accurate coverage of some bays, as is evident in the spatial representation of Vrboska (Fig.~\ref{fig:fig13}a, bottom).

Overall, the comparison suggests that the main difference between Meteo-HySEA and AdriSC-ADCIRC results is linked to the energy trapping within the semi-enclosed basins. While this behavior might reflect differences in dissipation schemes, friction parametrization, numerical approaches or horizontal resolutions, it also points to Meteo-HySEA’s capacity to capture the inundation and its impact on the resonant amplification within the bays. In particular, Figure \ref{fig:fig14} shows the maximum inundation computed at the three main locations using Meteo-HySEA. Significant flooding can be observed in Vela Luka and Stari Grad. In contrast, capturing inundation in Vrboska proves to be more challenging, as both 7 m spatial resolution and low-resolution topo-bathymetric data are insufficient to resolve the narrow channel characteristic of this area. 

\section{Concluding remarks}\label{sec:conclusions}

This study evaluated the GPU-based \textit{Meteo-HySEA} model as a new tool for meteotsunami simulation in the Adriatic Sea, assessing its performance against the CPU-based AdriSC-ADCIRC system and validating results with three well-documented historical events (2014, 2017, 2020). The main findings are as follows:

\begin{itemize}
\item Meteotsunami simulation accuracy is fundamentally limited by the quality of atmospheric forcing. Even with high-resolution WRF downscaling, mesoscale pressure disturbances are often underestimated in intensity and spatial variability, constraining the realism of modeled oscillations. Improved reanalyses, ensemble forecasts, and data assimilation tailored to meteotsunami-prone regions remain essential.

\item Meteo-HySEA generally reproduces the timing, spatial patterns, and overall structure of observed events, and in some cases produces stronger responses than AdriSC-ADCIRC under identical forcing. However, its tendency to overestimate dominant oscillation periods, particularly in semi-enclosed basins, suggests that harbor resonance and dissipation processes require refinement. This issue may stem from the wet–dry technique, in which the geomorphology of basins and harbors evolves over time as the wet areas are updated.

\item A major advantage of Meteo-HySEA lies in computational efficiency: GPU acceleration enables high-resolution multi-grid simulations to run orders of magnitude faster than conventional CPU-based models. This capability supports real-time operational use, ensemble forecasting, and explicit treatment of atmospheric and boundary-condition uncertainties.

\item Meteo-HySEA also directly simulates inundation processes, extending the modeling chain from offshore oscillations to onshore flooding. This functionality is particularly relevant for risk assessment and civil protection, as it allows the estimation of direct impacts on vulnerable harbors and urban waterfronts. In this regard, it is evident that high-quality, high-resolution topobathymetric data are required in the nearshore areas to fully exploit this core feature.
\end{itemize}

Overall, the results demonstrate that GPU-based solvers such as Meteo-HySEA represent a promising pathway toward next-generation meteotsunami forecasting and hazard assessment systems. At present, atmospheric forcing remains the dominant source of forecast uncertainty, but ongoing advances in high-resolution weather prediction—including convection-permitting ensembles and machine-learning-based nowcasting—are expected to improve forcing fidelity.

Future work should focus on two complementary directions. First, systematic validation with dense observational networks (tide gauges, pressure sensors, HF radar) is needed to determine whether the longer persistence of oscillations in Meteo-HySEA reflects more realistic harbor seiches or ADCIRC’s stronger damping better represents physical energy dissipation. Second, the operational potential of Meteo-HySEA should be tested through integration with real-time atmospheric forecasts and early warning protocols, ideally in collaboration with hydro-meteorological services and civil protection agencies.

From a broader perspective, GPU-accelerated inundation modeling tailored to meteotsunamis carries direct societal relevance. By reducing computational barriers, Meteo-HySEA enables rapid ensemble forecasts, probabilistic hazard assessments, and high-resolution inundation scenarios in near-real time. Such capabilities represent a step forward in anticipating meteotsunami impacts and providing actionable information to decision-makers, ultimately strengthening coastal resilience in the Adriatic

\backmatter
\bmhead{Acknowledgements}
Special thanks are given for the support of the European Centre for Medium-Range Weather Forecasts (ECMWF) staff and for the ECMWF computing and archive facilities used in this research.

\section*{Statements and Declarations}
\begin{itemize}
\item \textbf{Funding}. This work was funded by ChEESE-2P (EU-EuroHPC JU-101093038) and DT-GEO project (HORIZON-INFRA-2021-TECH-01-01, number 101058129). This research has also been funded by Red Española de Supercomputación (RES), grant AECT-2023-2-0033.
\item \textbf{Competing interests}. 
The authors have no relevant financial or non-financial interests to disclose.
\item \textbf{Author Contributions.}
All authors contributed to the study conception and design. Material preparation, data collection and analysis were performed by Alejandro González del Pino and Clea Lumina Denamiel. The first draft of the manuscript was written by Clea Lumina Denamiel and Alejandro González del Pino and all authors commented on previous versions of the manuscript. All authors read and approved the final manuscript.
\end{itemize}

\pagebreak

\begin{appendices}
\section{Atmospheric forcing evaluation}\label{secA}

\begin{figure}
    \centering
    \includegraphics[width=0.75\linewidth]{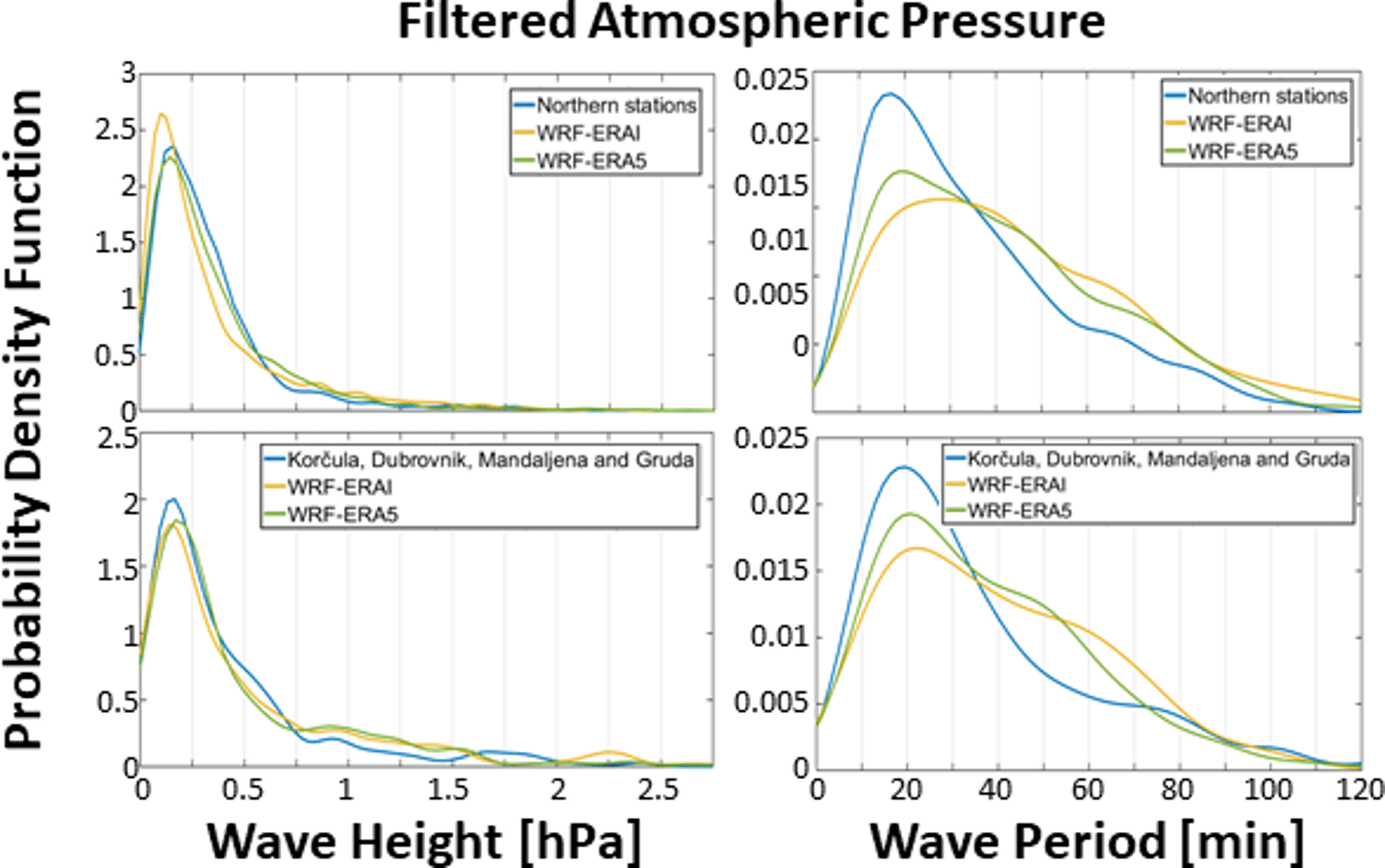}
    \caption{Meteotsunami events of June 25 and 26 2014. Probability density functions of the wave heights and periods extracted from the observed and modeled filtered atmospheric pressure at the Crometeo stations divided between the northern stations and the southern stations of Korčula, Dubrovnik, Gruda and Mandaljena.}
    \label{fig:figA1}
\end{figure}

\begin{figure}
    \centering
    \includegraphics[width=0.75\linewidth]{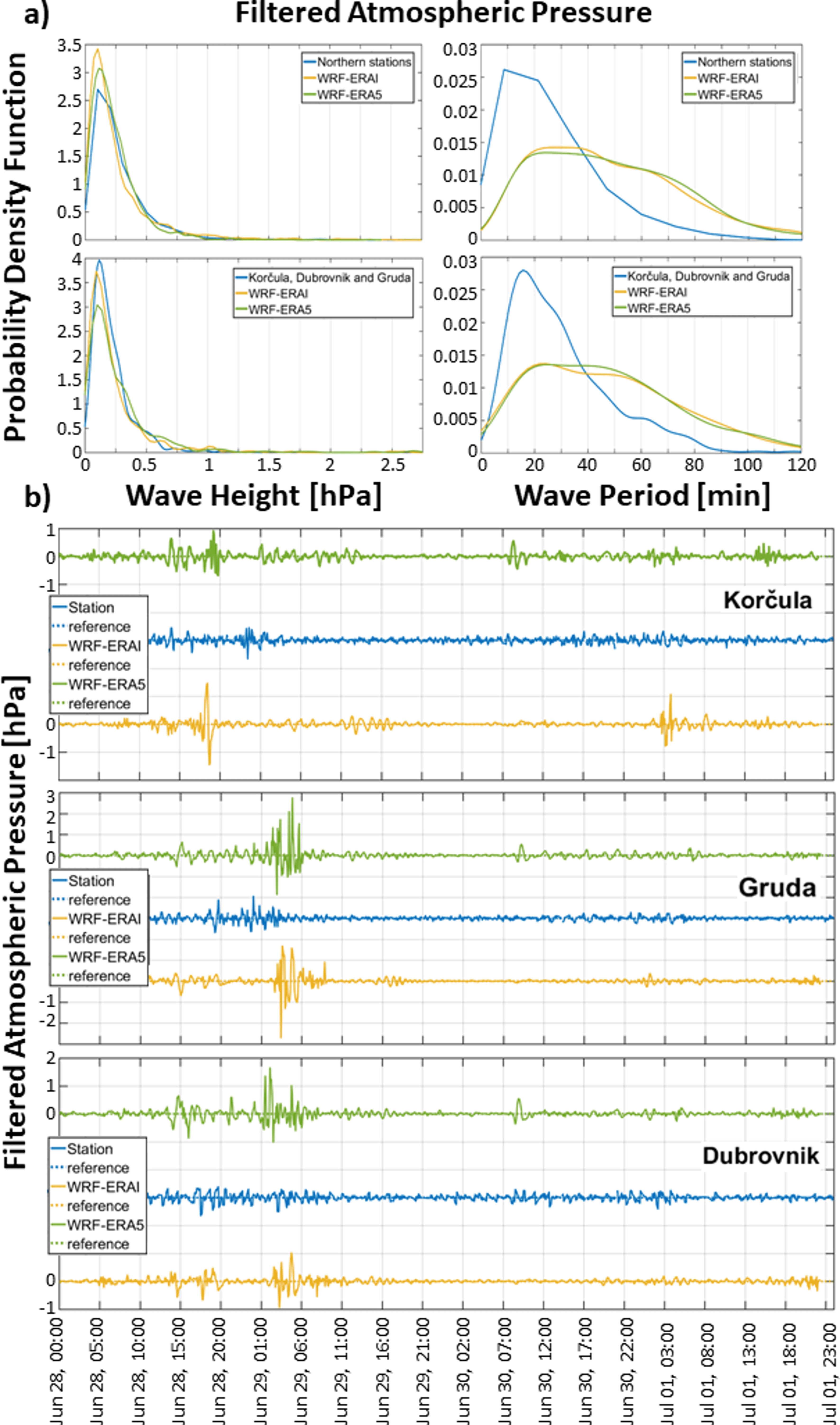}
    \caption{Meteotsunami events of June 28 to July 02 2017. (a) Probability density functions of the wave heights and periods extracted from the filtered atmospheric pressure at the Crometeo stations divided between the northern stations and the southern stations of Korčula, Dubrovnik and Gruda. (b) Observed and modeled filtered atmospheric pressure at the Korčula, Gruda and Dubrovnic Crometeo stations.}
    \label{fig:figA2}
\end{figure}

\begin{figure}
    \centering
    \includegraphics[width=0.75\linewidth]{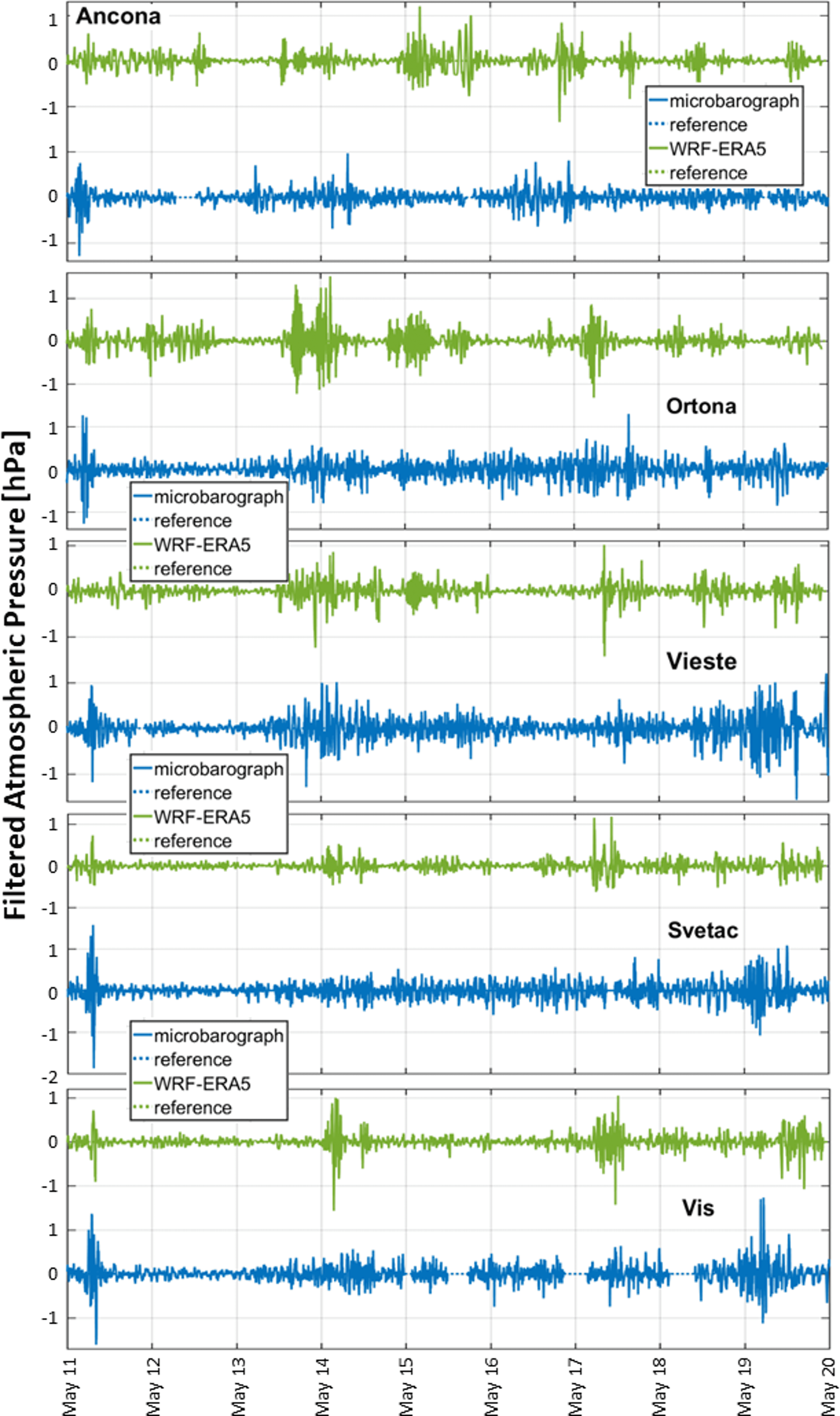}
    \caption{Meteotsunami events of May 11 to May 20 2020. Observed and modeled filtered atmospheric pressure at the Ancona, Ortona, Vieste, Svetac and Vis MESSI microbarograph locations.}
    \label{fig:figA3}
\end{figure}

In this annex we provide additional analyses of the atmospheric pressure disturbances associated with the studied meteotsunami events of June 2014, June-July 2017 and May 2020. Figure \ref{fig:figA1} shows the probability density functions of wave heights and periods derived from observed and modeled filtered atmospheric pressure at the Crometeo stations during the June 25–26, 2014 events, with results separated between northern stations and southern stations of Korčula, Dubrovnik, Gruda, and Mandaljena. Figure \ref{fig:figA2} presents similar analyses for the June 28–July 2, 2017 events: probability density functions of wave heights and periods at northern stations and the southern stations of Korčula, Dubrovnik, and Gruda and comparison of observed and modeled filtered atmospheric pressure at the Korčula, Gruda, and Dubrovnik stations. Finally, Figure \ref{fig:figA3} illustrates the May 11–20, 2020 events through comparisons of observed and modeled filtered atmospheric pressure recorded at the Ancona, Ortona, Vieste, Svetac, and Vis MESSI microbarograph locations.

\pagebreak

\section{Evaluation of Meteo-HySEA forced with WRF-ERAI}\label{secB}

\begin{figure}
    \centering
    \includegraphics[width=0.75\linewidth]{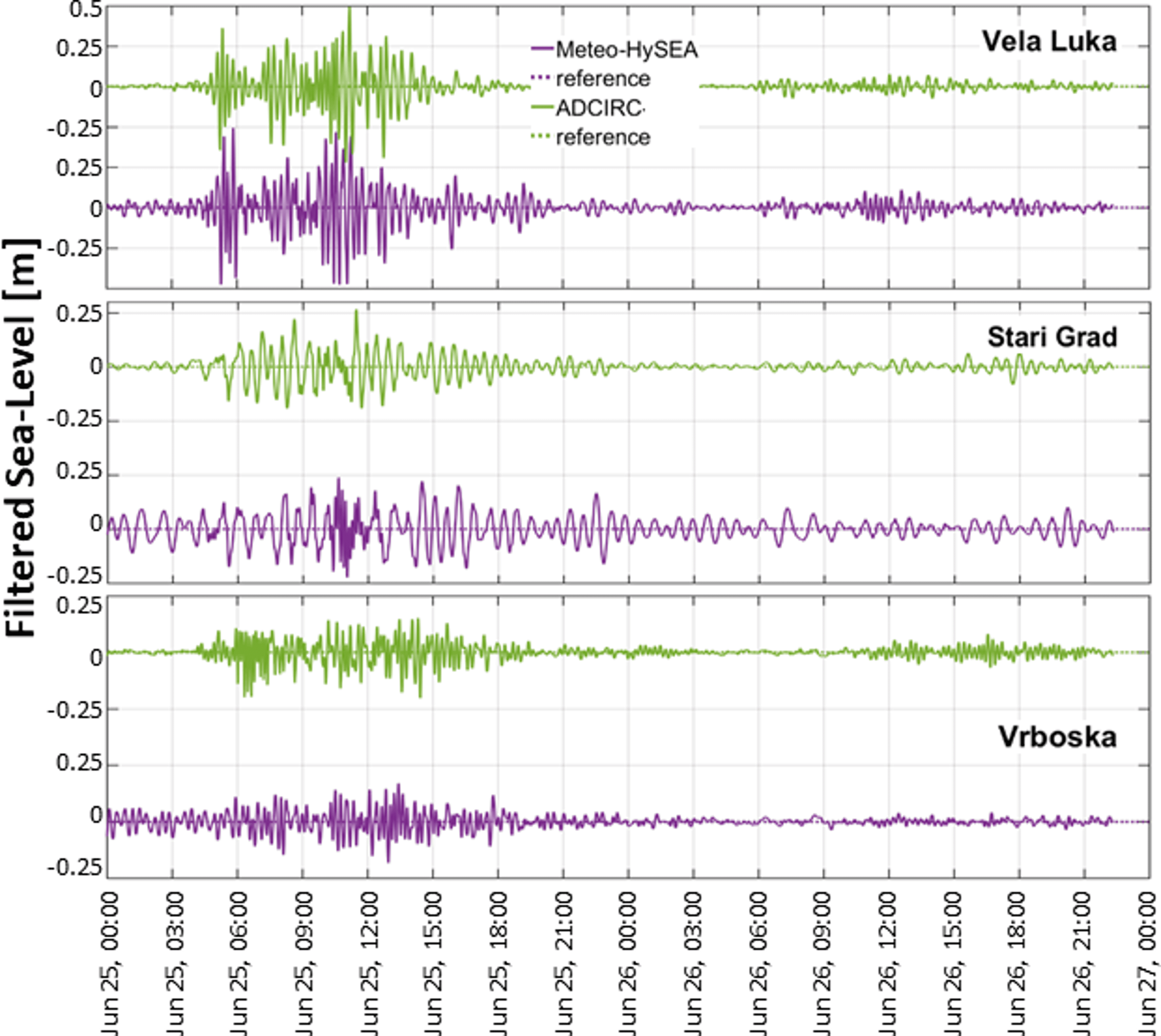}
    \caption{Meteotsunami events of June 25 and 26 2014 at Vela Luka, Stari Grad and Vrboska locations. }
    \label{fig:figB4}
\end{figure}

\begin{figure}
    \centering
    \includegraphics[width=0.75\linewidth]{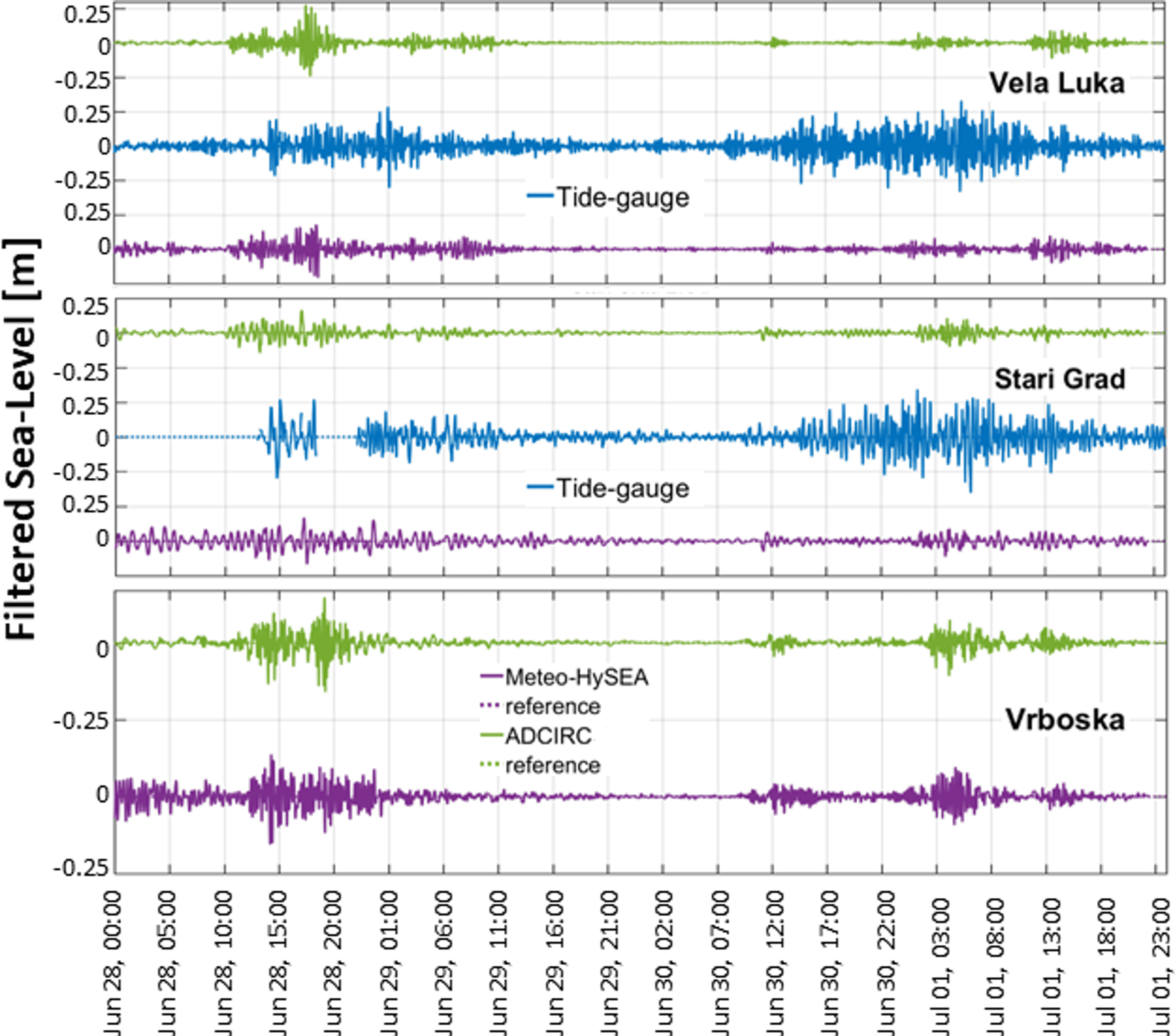}
    \caption{Meteotsunami events of June 28 to July 1 2017 at Vela Luka, Stari Grad and Vrboska locations. }
    \label{fig:figB5}
\end{figure}

In this annex we present an additional evaluation of Meteo-HySEA performance when forced with WRF-ERAI. Figure \ref{fig:figB4} illustrates the June 25–26, 2014 meteotsunami events at the Vela Luka, Stari Grad, and Vrboska locations, while Figure \ref{fig:figB5} shows the June 28–July 1, 2017 events at the same sites. These results provide a complementary assessment of model skill under alternative atmospheric forcing conditions.

Similarly to the results obtained with the WRF-ERA5 forcing (Figs. \ref{fig:fig07}a and \ref{fig:fig08}a), compared to AdriSC-ADCIRC, Meteo-HySEA reproduces stronger peak amplitudes in Vela Luka but weaker amplitudes in Stari Grad and Vrboska during the June 2014 events while, for the June–July 2017 events, both Meteo-HySEA and AdriSC-ADCIRC fail to reproduce the long-lasting oscillations observed at Vela Luka and Stari Grad despite capturing some oscillations on both June 28 and July 1.

\end{appendices}

\pagebreak








\end{document}